\documentclass[aps,pre, twocolumn, showpacs,groupedaddress]{revtex4}

\usepackage{times}
\usepackage{graphicx}
\usepackage{amsmath,amsfonts}
\newcommand{\qmax}{q_{\text{max}}}
\newcommand{\mr}{m({\bf r})}
\newcommand{\vmr}{{\bf m(r)}}

\newcommand{\fcr}{f^c({\bf r})}
\newcommand{\fCr}{f^C({\bf r})}
\newcommand{\ve}{ \mathbf{e} }
\newcommand{\ii}{\mathrm{i} }
\newcommand{\ee}{\mathrm{e}}
\newcommand{\muT}{\mu_{\text{tot}}}

\newcommand{\eeqref}[1]{eq. \eqref{#1}}

\makeatletter
\newif\if@restonecol
\makeatother

\begin{document}

\title{Recovering magnetization distributions from their noisy diffraction data} 

\author{Ne-Te Duane Loh$^{1,\,2}$}
\author{Stefan Eisebitt$^{3}$}
\author{Samuel Flewett$^{3}$}
\author{Veit Elser$^1$}
\affiliation{$^1$ Laboratory of Atomic and Solid State Physics (LASSP) Cornell University, Ithaca, NY 14853-2501, USA\\
$^2$ Cornell High Energy Synchrotron Science (CHESS) Cornell University, Ithaca, NY 14853-2501, USA\\
$^3$ Institut f\"{u}r Optik und Atomare Physik, TU Berlin, Stra\ss e des 17. Juni 135, D-10623 Berlin, Germany} 

\date{\today}

\begin{abstract}
We study, using simulated experiments inspired by thin film magnetic domain patterns, the feasibility of phase retrieval in X-ray diffractive imaging in the presence of intrinsic charge scattering given only photon-shot-noise limited diffraction data. We detail a reconstruction algorithm to recover the sample's magnetization distribution under such conditions, and compare its performance with that of Fourier transform holography. Concerning the design of future experiments, we also chart out the reconstruction limits of diffractive imaging when photon-shot-noise and the intensity of charge scattering noise are independently varied. This work is directly relevant to the time-resolved imaging of magnetic dynamics using coherent and ultrafast radiation from X-ray free electron lasers and also to broader classes of diffractive imaging experiments which suffer noisy data, missing data or both.   
\end{abstract}


\maketitle

\section{Introduction.}

There has been a growing interest in studying and manipulating thin-film magnetic nanostructures \cite{Elser:2010, Saga:1999, Eisebitt:2003, Eisebitt:2004, Pierce:2003, Pierce:2007, Gutt:2010}. Besides the commercial applicability of such studies, experimental data on the formation, dynamics and stability of magnetic nanostructures will provide clues for constructing predictive models of magnetic materials \cite{Pierce:2007}, which may in turn drive the invention of novel devices. 

\begin{figure}[t]
\centering
\includegraphics[width=1.6in]{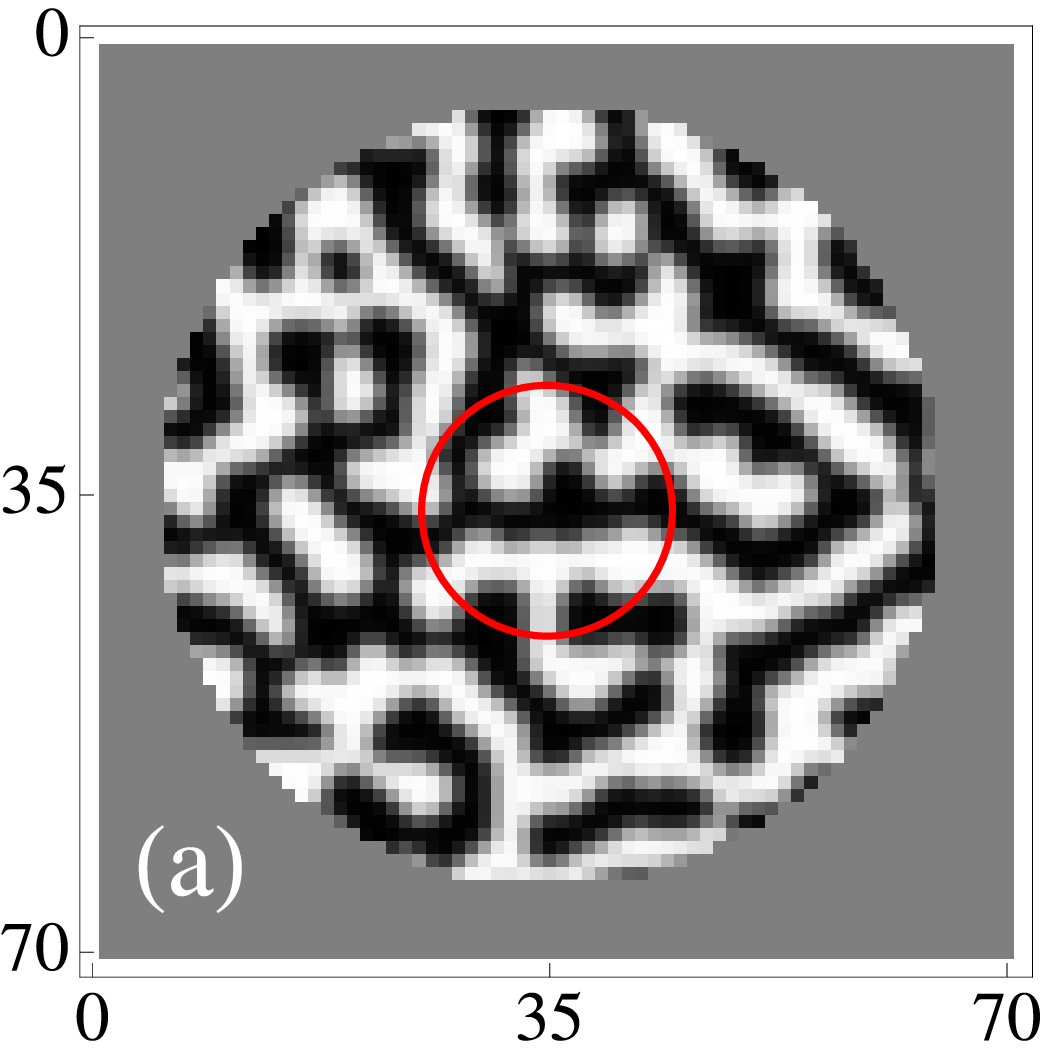}
\includegraphics[width=1.6in]{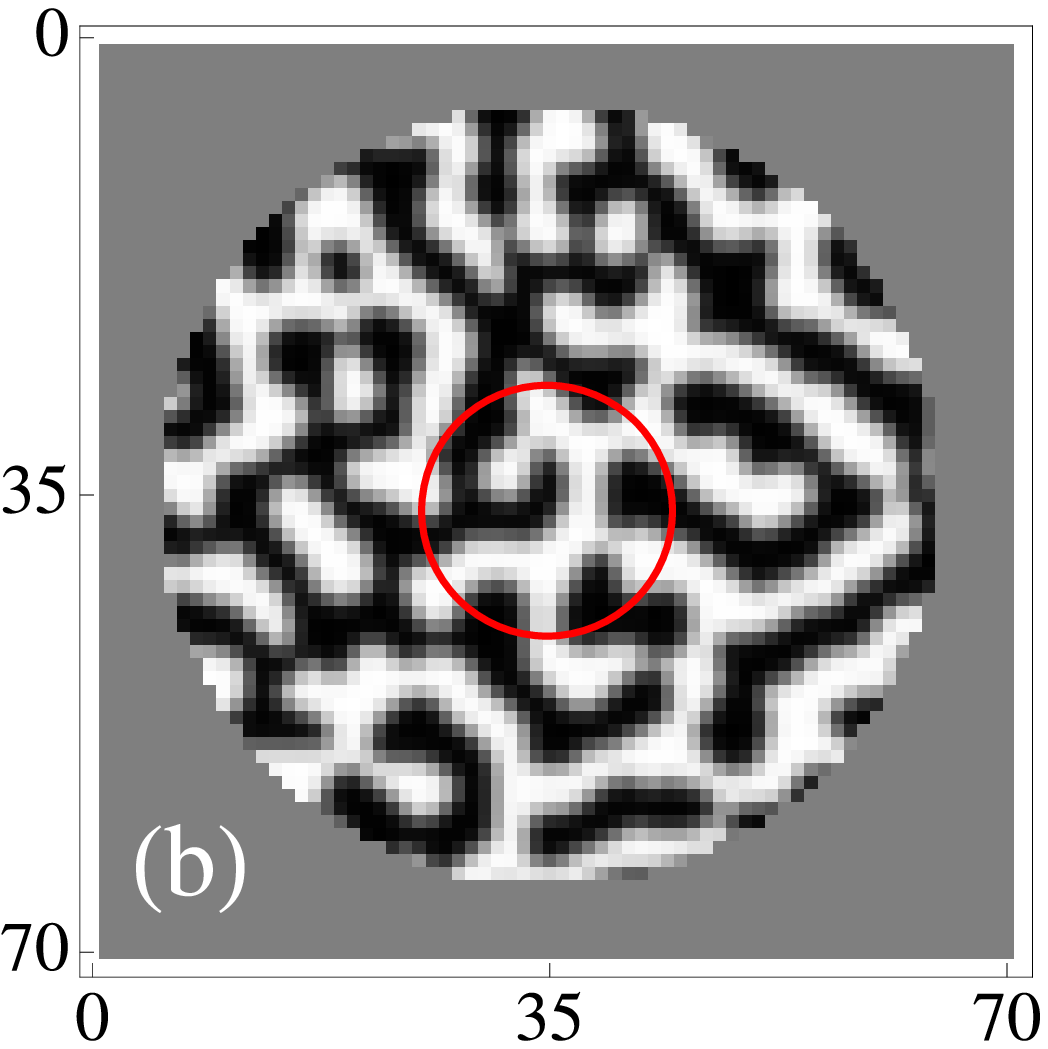}
\caption{Simulated example of the type of magnetic domain motion that could be investigated by an XFEL source. The magnetization distributions above differ only within the central circle. The average domain is 5 pixels wide, or approximately 170 nm when related to the experiment data in \cite{Eisebitt:2004} (each pixel hence measures 34 nm). We explain how to generate an ensemble of such similar distributions in the text.}\label{fig:Dynamics}
\end{figure}

A comprehensive understanding of these magnetic nanostructures involves studying extremely fast magnetic dynamics at high resolution. Ideally, this can be achieved by sequentially illuminating an evolving magnetic specimen using very short, intense pulses of coherent X-ray radiation (image of such in Fig. \ref{fig:Dynamics}). Such radiation has become available at X-ray free electron laser (XFEL) facilities, which can produce femtosecond pulses with upwards of $10^{12}$ X-ray photons each. Despite such high intensities, pulses are typically monochromatized and polarized for magnetic imaging at the expense of their photon flux. Furthermore, in the case of repetitive studies on the same sample, say to study a specimen's dynamics, the intensity of the XFEL pulses may have to be reduced to prevent sample damage by energetic X-ray photons. As a result of reducing pulse intensity, the diffraction signal from the weakly scattering magnetic contrast in specimens are often expected to be {\it photon-shot-noise} limited \cite{Gutt:2010}. To make matters worse, the magnetic signal may also be contaminated by strong scattering from the non-uniform charge density intrinsic to magnetic specimens.

Currently, Fourier transform holography \cite{Eisebitt:2004} and speckle metrology \cite{Pierce:2003} are two leading coherent X-ray techniques already used to study magnetic nanostructures. Their effectiveness comes with limitations: speckle metrology is restricted to ``fingerprinting'' in reciprocal space (unable to resolve localized dynamics of Fig. \ref{fig:Dynamics}); Fourier transform holography affords direct-space imaging but it requires the crafting of a reference structure. 

This paper describes an alternative to Fourier transform holography and speckle metrology, detailing a reconstruction algorithm that directly images extended magnetization distributions when given only transmission diffraction data, without the need for a reference illumination. This algorithm crucially uses prior information about the ensemble of magnetization distributions to reconstruct a specific distribution. To demonstrate this, we had to generate an ensemble of credible magnetization distribution to be used as scattering sources for our diffractive imaging simulations (Section \ref{sec:GenerateDomains}). Whereas our algorithm is robust when the ensemble magnetization values are known, it is still relevant even when given limited information about these values (Appendix \ref{appendix:SimpleDirectSpaceConstraint}). We expect our algorithm to apply, with possibly reduced efficacy, to real magnetization distributions whose ensemble properties are less well characterized but qualitatively similar to those in this paper. 

Our algorithm also exploits knowledge of the sample's direct-space support to reduce the effects of photon-shot-noise in the diffraction data. To make our demonstrations relevant to ultrafast magnetic imaging with XFEL radiation, we use diffraction data with the severest levels of noise in addition to missing data in the beamstop. We compare our diffractive imaging reconstruction with the performance of Fourier transform holography at comparable noise levels. 

To substantiate our methods, we include a feasibility study of noisy magnetic imaging when subjected to our described methods in the most optimistic scenario that the ensemble magnetization values are known. We expect this feasibility study to be useful in the design of future experiments.

\section{Resonant magnetic scattering.}

Multilayer magnetic thin films with perpendicular magnetic anisotropy \cite{Eisebitt:2003, Pierce:2003, Eisebitt:2004, Pierce:2007, Gutt:2010, Hannon:1988, Kortright:2001} exhibit a notable phase comprising magnetic nanostructures that can be described by a 2D coarse-grained magnetization distribution $\vmr$. The magnetization in this phase is effectively parallel or antiparallel to the sample's layer normal \footnote{Depending on the magnetostatic and domain wall energy, closure domains with in-plane magnetization may form where the domain walls meet the film surface \cite{Durr:1997}. These closure domains are negligible in the very thin films which we examine in this paper.}. In this section, we briefly discuss how such magnetization distributions are encoded in the diffraction data.  

In diffractive imaging, one typically measures the sample's elastic photon scattering amplitude which varies across the sample. This scattering mechanism includes virtual transitions between core electron states and unoccupied electron states above the Fermi level \cite{Hannon:1988}. Since these unoccupied states are spin-polarized due to the sample's local magnetization, the photon scattering amplitude depends on the sample's magnetization distribution $\vmr$. 

There are, naturally, other components of the sample's elastic scattering amplitude that are insensitive to the magnetization: $f^0({\bf r})$, the Thomson contribution; $\fcr$, the anomalous charge scattering. Both of these contributions are integrated along the incident beam direction. Like $\vmr$, $f^0({\bf r})$ and $\fcr$ are also treated as 2D distributions. 

A magnetic specimen's total elastic scattering amplitude is given by \cite{Hannon:1988} as
\begin{eqnarray}
f_{\text{tot}} ({\bf r}) &\approx& f^0({\bf r}) + ({\ve_{f}}^{\ast} \cdot  \ve_{i}) \,\fcr - \mathrm{i} ({\ve_{f}}^{\ast} \times \ve_{i}) \cdot \vmr \, f^M  \nonumber \\
& & + \left({\ve_{f}}^{\ast} \cdot \vmr \right) \left({\ve_{i}}^{\ast} \cdot \vmr \right) \, f^{m}\label{eqn:CrossSection} \; ,
\end{eqnarray}
where $\ve_{i}$ and $\ve_{f}$ are the polarization vectors of the incident and scattered radiation. The magnetization-sensitive scattering amplitudes $f^M$ and $f^m$ are scaled to allow the magnetization to be normalized as $\text{max}(|\vmr|) = 1$.

The total elastic scattering amplitude $f_{\text{tot}} ({\bf r})$ of multilayer magnetic films can be simplified with a few experimental constraints. First, since the magnetization is parallel or antiparallel to the sample's layer normal, we can replace $\vmr$ with the longitudinal scalar distribution $\mr$. More importantly, the contribution to the scattering amplitude from the last term in \eeqref{eqn:CrossSection} is suppressed if light were transmitted along this longitudinal direction. Second, we restrict ourselves to circularly polarized incident radiation, which is a scattering eigenstate of the 3rd term in \eeqref{eqn:CrossSection}. This choice, however, causes the diffraction patterns from magnetic and charge distributions to interfere \footnote{There will be no interference between charge and magnetic scattering terms if the incident radiation were linearly polarized. In this case, diffraction intensities from charge and magnetic distributions are separately added, as demonstrated in reference \cite{Eisebitt:2003}, and the former, ideally, can be subtracted away. Determining the static, random charge scattering for subtraction is possible when the photon energy is detuned away from the core-level resonance, hence suppressing magnetic scattering. But this subtraction may be unreliable at noisy, high-$q$ signal regions where the magnetization distribution is primarily encoded. Subtraction might also be problematic in single-shot imaging, when the incident photon fluence may fluctuate between shots --- guesswork is needed to match the intensities of the charge-plus-magnetic data to those of charge-only data for reliable subtraction.}. Third, in the small-angle scattering limit, we can combine the non-magnetic scattering contributions as $f^C({\bf r})$. These conditions produce a simplified total scattering amplitude:
\begin{equation}
f_{\text{tot}}({\bf r}) \approx \fCr + f^M \mr \; \label{eqn:SimpleCrossSection} \; .
\end{equation}
Experimently, the magnetic scattering amplitude $f^M$ can be dramatically increased through resonant scattering: by tuning the energy of the incident photons to match those of core-level electron transitions in the sample (L or M edges) \cite{Eisebitt:2003, Pierce:2003, Eisebitt:2004, Pierce:2007, Gutt:2010, Hannon:1988, Kortright:2001}. This enhances the scattering signature of the magnetization with respect to the charge distribution, which is useful since we are interested only in the former. 

A difference in the correlation length of the charge distribution and that of the magnetization distribution is common, which causes a separation in the peaks of their respective diffracted power \cite{Kortright:2001}. Potentially, one could then ignore the charge distribution when imaging the magnetization at a lower resolution. However, later sections of this paper show that magnetic imaging may still be difficult despite such a separation. 

The Fraunhofer diffraction intensity from samples obeying \eeqref{eqn:SimpleCrossSection} is
\begin{equation}
I({\bf q}) \propto \phi \left| \int \text{d}^2 {\bf r}\; \ee^{\ii {\bf q} \cdot {\bf r}}\, f_{\text{tot}}({\bf r}) \right|^2 \label{eqn:ScatteredIntensity}\; ,
\end{equation}
where $\phi$, the photon fluence, crucially determines the number of diffracted photons and hence the severity of photon-shot-noise. Since only the total number of diffracted photons in our simulations is experimentally relevant and can be varied by changing only $\phi$, the absolute scale of the magnetic and charge scattering amplitudes in $f_{\text{tot}}({\bf r})$, $f^M$ and $f^C$ respectively, is immaterial. From here on, magnetic scattering amplitudes and magnetization become interchangeable because they differ only by this unimportant absolute scale. The same is true between charge scattering amplitudes and charge. Since it still serves to be consistent, we normalize the magnetic scattering amplitude $f^M = 1$ in \eeqref{eqn:SimpleCrossSection}. The ratio $f^C/f^M$, however, depends on the experimental specimen and the polarization of the incident radiation. This means, of course, that $\phi$ is no longer strictly the photon fluence, but a variable to control the number of scattered photons.    

\section{Generating magnetic domain patterns.}\label{sec:GenerateDomains}

\begin{figure}[t!]
\centering
\includegraphics[width=0.5\textwidth]{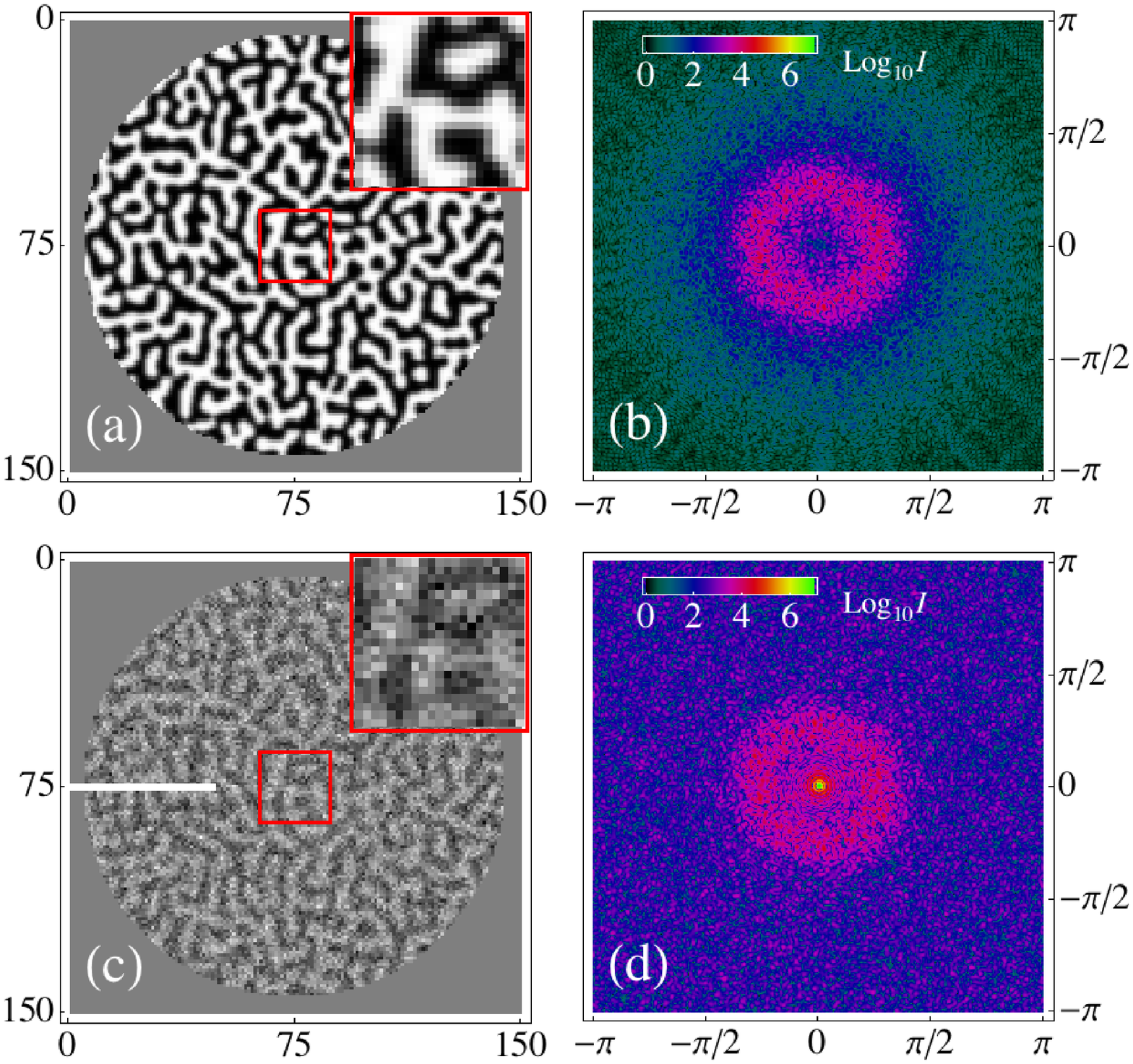}
\includegraphics[width=0.47\textwidth]{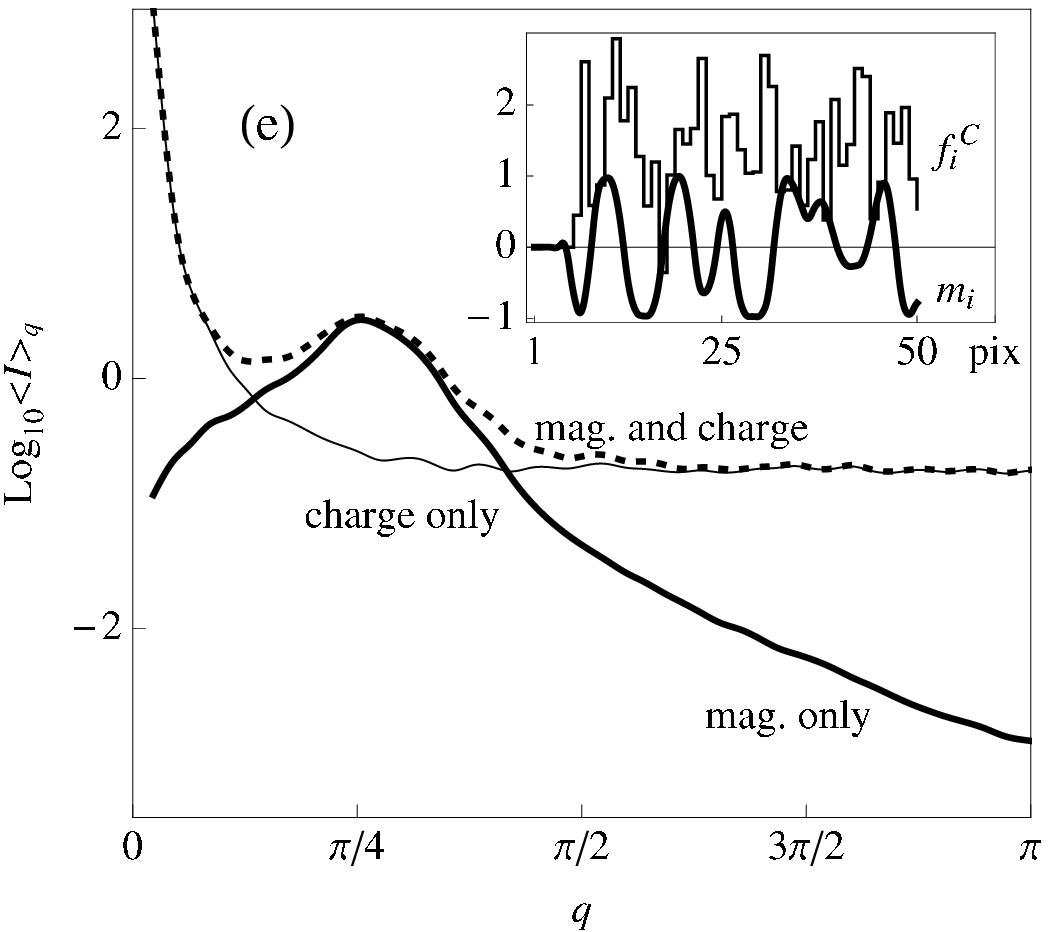}
\caption{(color online). Simulated magnetic domains and their diffraction intensities, which together illustrate the effects of charge scattering. Panel {\bf (a)} shows domains without charge scattering and in {\bf (b)} the logarithm of their diffraction intensities. In {\bf (c)} the same domains with random charge distribution ($\Delta_m/\Delta_c = 1$) added and {\bf (d)}, the logarithm of its resultant diffraction intensities. Panel {\bf (e)} plots the azimuthally averaged diffraction intensities from the charge only (thin, solid line), magnetic only (thick, solid line) and charge-plus-magnetization distributions (dashed line) belonging to the domain pattern in (c). The inset in (e) plots the scattering amplitudes ($f^C_i$ for charge; $m_i$ for magnetic) along the horizontal white line of the simulated domain pattern in (c). }\label{fig:MagAndCharge}
\end{figure}

To simulate realistic magnetic imaging, we first need to generate magnetization distributions, or {\it domain patterns}, that resemble a wide and interesting variety of actual specimens. At a minimum, the ensemble of such domain patterns should conform to these experimental observations: 
\begin{enumerate}
\item{} in Fourier-space, an azimuthally symmetric diffracted power which peaks at a particular spatial frequency (compare simulated example in Fig. \ref{fig:MagAndCharge}b to those from experiments in \cite{Kortright:2001, Eisebitt:2003, Eisebitt:2004, Pierce:2007, Gutt:2010}); 
\item{} in direct-space, a statistical distribution on the magnetization (Fig. \ref{fig:BinaryContrast}) of ferromagnetic domains with finite-width domain walls (Fig. \ref{fig:MagAndCharge}a).
\end{enumerate}

The clues to generating realistic domain patterns lie in the careful examination of the diffraction envelope shown in Fig. \ref{fig:MagAndCharge}e. The spatial frequency dependence of this envelope reveals two competing effects: short-range exchange interaction that produces ferromagnetic domains and long-range demagnetizing fields which in turn destabilize these domains.

These effects are modeled by the 2D Landau-Ginzburg free energy density 
\begin{eqnarray}\label{eqn:LG}
\mathcal{F}(\mathbf{r})&=& A\left(m(\mathbf{r})^2-1\right)^2+B\,|\nabla m(\mathbf{r})|^2+ \nonumber \\
& & C\int_{|\mathbf{r}'-\mathbf{r}|>l}\frac{m(\mathbf{r})m(\mathbf{r}')}{|\mathbf{r}-\mathbf{r}'|^3}\,d^2\mathbf{r}'\; ,
\end{eqnarray}
where $A$, $B$ and $C$ are temperature dependent positive quantities, and $l=\pi/q_\mathrm{max}$ is a cutoff that defines the maximum spatial frequency. Rewriting (\ref{eqn:LG}) in terms of the Fourier modes of the magnetization, $m(\mathbf{q})$, we obtain in the limit $|\mathbf{q}|=q\ll q_\mathrm{max}$ the following expression:
\begin{eqnarray}
\mathcal{F}(\mathbf{q})&=& |m(\mathbf{q})|^2 
\left[ 
\begin{array}{ll} -2A+B q^2+ \\
\\
2\pi C\left( q_\mathrm{max}/\pi -q+ (\pi/4)\, q^2/q_\mathrm{max} \right) \end{array}
\right] \nonumber \\ 
&+&O\left(|m(\mathbf{q})|^4\right).
\end{eqnarray}
Defining new constants $a>0$ and $b$, and rescaling $q_\mathrm{max}$ by a constant, this can be rewritten in the much simplified form
\begin{equation}
\mathcal{F}(\mathbf{q})=\left[(q/q_\mathrm{max}-a)^2+b\right]|m(\mathbf{q})|^2+O\left(|m(\mathbf{q})|^4\right).
\end{equation}
The ferromagnetic instability corresponds to $b\to 0$ and the $q$-dependence of the fluctuations as this limit is approached is controlled by the coefficient of the term quadratic in the magnetization. As a simple model for the formation of magnetic domains in real materials we will assume the distribution of fluctuations in the paramagnetic phase ($b>0$), given by the equipartition theorem, is preserved when the system freezes into a particular domain pattern. The intensity in this model is given by
\begin{equation}
I(q)\propto \frac{1}{(q/q_\mathrm{max}-a)^2+b} \; .\label{eqn:IntensProfile}
\end{equation}
Our simulations will use this form for the power spectrum with $a$ and $b$ fitted to agree with experimental data in \cite{Eisebitt:2004}. We use dimensionless units where the maximum spatial frequency $q_\mathrm{max}$ is scaled to the value $\pi$.

The generation of each domain pattern begins with an array of random, uniformly distributed real numbers between -1 and +1, mimicking the high-temperature magnetization distribution in the absence of external fields. On this random state $\mr$, we apply two nonphysical operations in turn: 
\begin{enumerate} 
\item{} band-pass Fourier filter using \eeqref{eqn:IntensProfile} --- 
\begin{equation}
m({\bf q})  \to  \frac{m({\bf q}) }{\sqrt{(q/\qmax - 0.27)^2 - 0.015}} \label{eqn:Lorentzian}  \;,
\end{equation}
where $m({\bf q})$ is the discrete Fourier transform of $\mr$;
\item{} binary projection on $\mr$ --- 
\begin{equation}
\mr \to 
\begin{cases}
+1, & \mbox{if }\mr \geq 0\\
-1, & \mbox{if } \mr < 0
\end{cases} \; .
\end{equation}\label{eqn:BinaryProjection}
\end{enumerate} 
The composition of these two operations is iterated on $\mr$ until it {\it converges}, where the values of $\mr$ are unchanged upon further iteration. Thereafter, we simulated finite domain wall widths by multiplying the converged distribution $\mr$ with a final low-pass Fourier filter, $\exp{(-2.5 \,(q/\qmax)^2)}$ \footnote{To minimize the finite-size effect from sampling the distribution on a numerical array, we assumed that the coarse-graining length is considerably smaller than the width represented by one array pixel. We generated domains with twice the resolution ($|q_x| \leq 2 \qmax$ and $|q_y| \leq 2 \qmax$ without changing $\qmax$ in \eeqref{eqn:Lorentzian}) then truncated the Fourier-space of the converged domain pattern back to the lower resolution $\qmax$.}. This domain pattern is then normalized to $\text{max}(|\mr|) = 1$. 

Different, random initial arrays result in different domain patterns $\mr$, defining an ensemble of simulated patterns. Whereas we generated domains with zero net magnetization, this recipe can be easily modified to change this net magnetization. 

This recipe for generating domain patterns is easily extended to create perturbed versions of any domain pattern: we replace randomly selected circular areas in a previously converged source domain pattern with random numbers, then reapply the domain generation recipe until this perturbed pattern converges. This replacement occurs before the low-pass Fourier filter is applied to the source pattern. As an example, the pattern in Fig. \ref{fig:Dynamics}b is a converged perturbation of the pattern in Fig. \ref{fig:Dynamics}a. 

\section{Model of charge scattering.}

Since it is reasonable to expect the charge distribution $\fCr$ to be spatially uncorrelated at the resolution of the resonant scattering experiments \cite{Kortright:2001, Gutt:2010}, we model it as a 2D array of random, real numbers $\fCr$. Each array element of $\fCr$ represents the charge scattering amplitude averaged over a pixel.

The statistics of the spatially uncorrelated charge distributions is characterized by its mean $\langle \fCr \rangle$ and standard deviation, 
\begin{equation}
\Delta_c = \sqrt {\langle \, (\fCr - \langle \fCr \rangle)^2 \, \rangle } \label{eqn:DeltaC}\;, 
\end{equation}
which we coin {\it charge contrast}. The angle brackets denote the average over each distribution. The charge contrast should be compared to the {\it magnetic contrast}, 
\begin{equation}
\Delta_m = \sqrt {\langle \, (\mr - \langle \mr \rangle)^2 \, \rangle }  \label{eqn:DeltaM} \;. 
\end{equation}

The diffraction intensity in \eeqref{eqn:ScatteredIntensity} does not distinguish between charge and magnetic scattering, so any reconstruction can only determine their sum (see \eeqref{eqn:SimpleCrossSection}). Since we are interested only in recovering the magnetization distribution, the intrinsic charge distribution will contribute an inextricable scattering noise, characterized only by the signal-to-noise ratio $\Delta_m / \Delta_c$. When $\Delta_m  \approx \Delta_c$, it becomes visually impossible to differentiate between these distributions even if their sum were correctly reconstructed (compare Fig. \ref{fig:MagAndCharge}a and \ref{fig:MagAndCharge}c). 

In contrast, the mean charge scattering amplitude $\langle \fCr \rangle$, as later sections will show, is an immaterial constant to the reconstruction of $\mr$.  Nevertheless, to be consistent, we fixed $\langle \fCr \rangle = 1.33$ using experimental data from \cite{Eisebitt:2004}.

\section{Diffractive imaging as constraint satisfaction.}

We can interpret the diffractive imaging experiments in the language of constraint-satisfaction problems. Essentially, the goal of diffractive imaging is to recover the true magnetization distribution subject to two constraints: its measured noisy diffraction data (Fourier constraint) and the assumed statistics on its expected magnetization (direct-space constraint). This section discusses how we generated and characterized these two constraints.

\begin{figure}[t!]
\centering
\includegraphics[width=3.2in]{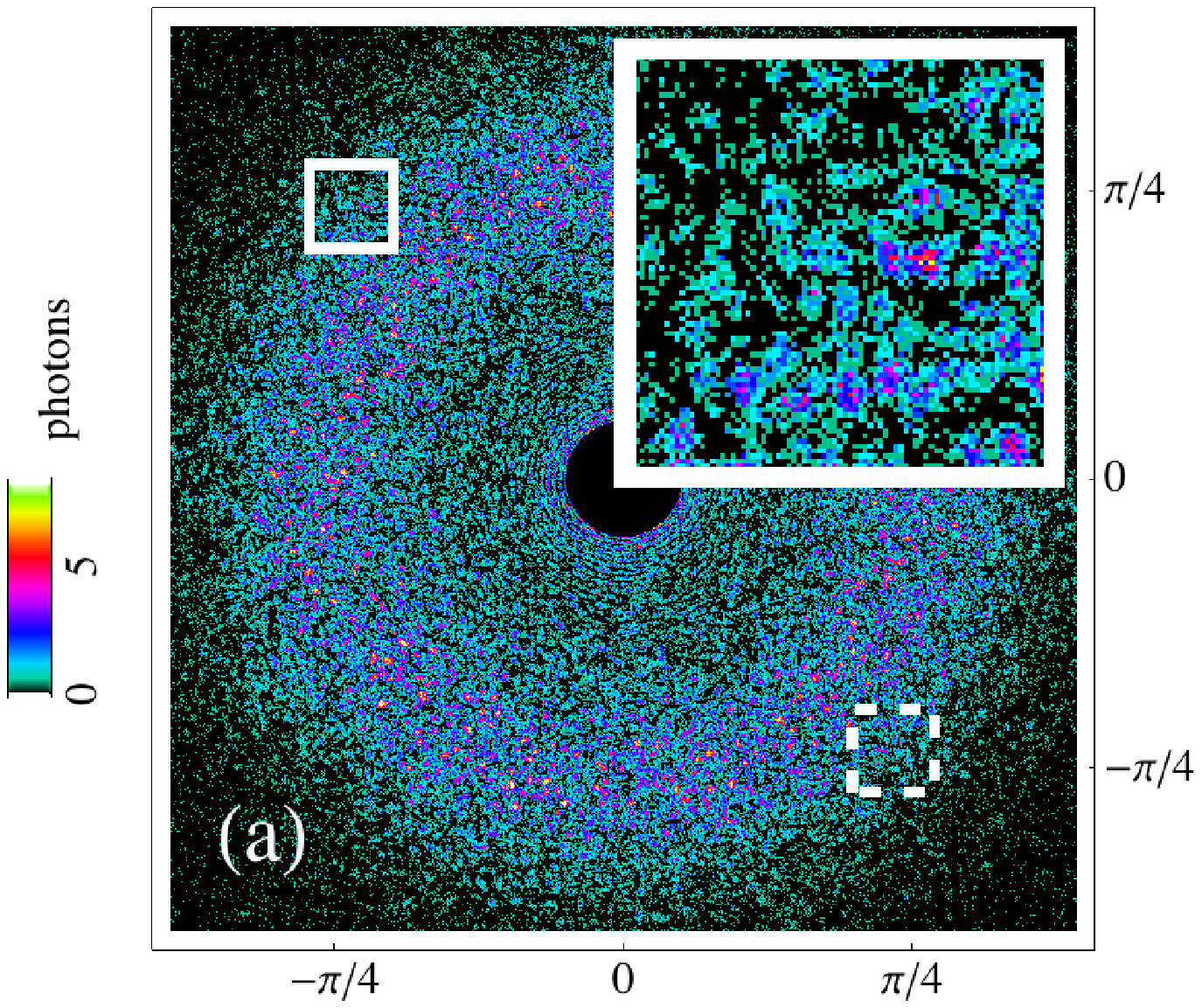}
\includegraphics[width=3.3in]{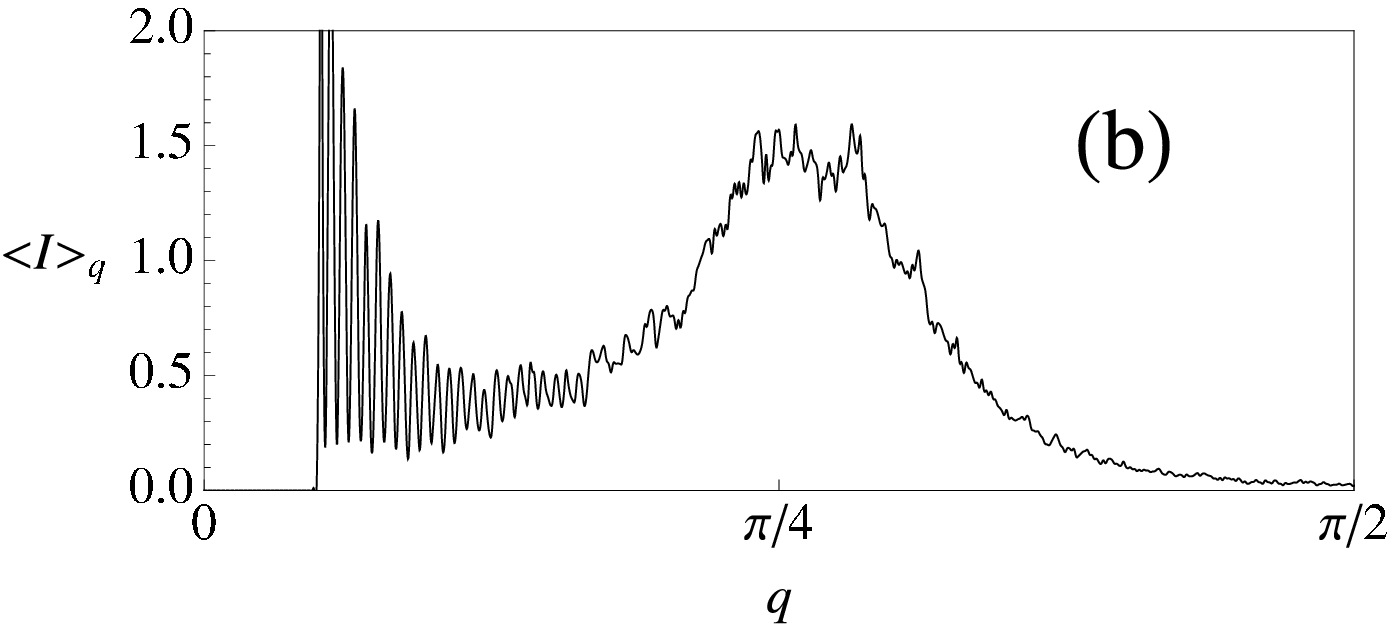}
\caption{(color online). Noisy photon data from simulated diffraction experiments. Panel {\bf (a)} shows the diffraction data from the domain pattern in Fig. \ref{fig:LargeRecon}a, with signal-to-noise of point B in Fig. \ref{fig:ContourPlot}. The intensities in the dashed-line box (lower right) of panel (a) are inversion symmetric to those in the solid-line box (upper left); larger inset is a magnified view of photon data in the solid-line box; the central black disk is the beamstop. Panel {\bf (b)} shows the azimuthally averaged photon counts in panel (a).}\label{fig:Data}
\end{figure}

\subsection{Fourier constraint.}

The Fourier constraint requires that the diffraction intensities of the true magnetization distribution, which we wish to recover, be statistically compatible with the measured photon data, mindful that the data includes intrinsic charge scattering. 

To simulate the diffraction data, we first added each pair of randomly-generated magnetization and charge distributions, $\mr$ and $\fCr$ respectively. We confined this total scattering amplitude to a circular support $S$ (Fig. \ref{fig:MagAndCharge}a, for example). Its continuous intensity distribution was scaled by $\phi$ to give the desired average number of scattered photons, then Poisson sampled to simulate photon-shot-noise. Following this, we averaged each data with its Friedel-symmetry counterpart to make it consistent with the real-valued direct-space contrast. Finally, a beamstop was applied to this symmetrized data, thus removing photon counts that would be contaminated by intense, unscattered radiation in actual experiments. The size of the beamstop was adjusted such that the remaining photon counts span two orders of magnitude (example photon data in Fig. \ref{fig:Data}). Naturally, the unmeasured Fourier amplitudes at spatial frequencies within the beamstop are unconstrained in our reconstruction algorithm. 

\begin{figure}[t!]
\centering
\includegraphics[width=3.3in]{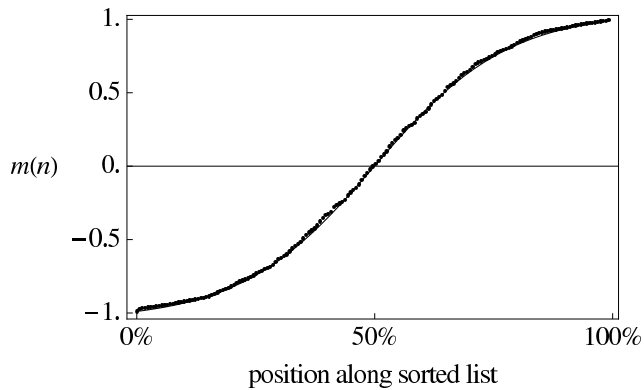}
\caption{Sorted-value magnetization constraint. Sorted list of normalized magnetization $\widetilde{m}(n)$ when averaged over many simulated domain patterns (curve) and those from one pattern (dots).  }\label{fig:BinaryContrast}
\end{figure}

\subsection{Direct-space constraint.}

When discussing the direct-space constraints on the magnetization it is convenient to introduce the sorted-value representation $m(n)$, where $m(1)$ is the smallest magnetization among the pixels within the support, $m(2)$ is the next smallest, etc. and $m(N_S)$ is the largest magnetization value. Within the ensemble of random domain patterns produced by the same magnetic material (and identical external parameters) the plots of the functions $m(n)$, with $n$ ranging from $1$ to $N_S$, should be nearly the same. Figure \ref{fig:BinaryContrast} compares $m(n)$ for one simulated domain pattern with the averaged $m(n)$ over many patterns. The structure of $m(n)$ is mainly a function of two lengths: the width of domains and the width of domain walls; materials with very thin domain walls will have a more step-like $m(n)$ \footnote{In thin magnetic films (multilayers with perpendicular anisotropy), the domain wall width scales as $\sqrt{A_s/K_u}$, where $A_s$ is the exchange stiffness constant and $K_u$ is the uniaxial anisotropy constant, both of which are constants of the material being probed \cite{Hubert}. One also finds that the domain width is proportional to the film thickness, and inversely proportional to the applied field \cite{Hubert}. As a result, a variety of contrast histograms may be observed depending on the specific material properties and geometry of the system under investigation. Here we use the magnetization constraint function shown in Fig. \ref{fig:BinaryContrast} as a prototypical example. In a real experiment, this constraint could be determined by calculating the expected widths of the domains and their walls using magnetic domain theory presented for example in \cite{Hubert} but properly relaxed to include intrinsic blurring in experimental diffractive imaging.}. 

The simplicity of the function $m(n)$ serves as a powerful constraint for the domain reconstructions. We will use the notation $\widetilde{m}(n)$ for the sorted-value magnetization constraint. In most of our simulations we will assume $\widetilde{m}(n)$ is known. At zero net magnetization we have the symmetry

\begin{equation}\label{eqn:MagSymmetry}
m(n) = -m(N_S+1-n) \;.
\end{equation}

Appendix \ref{appendix:SimpleDirectSpaceConstraint} briefly describes reconstructions that only use this property and boundedness of $m(n)$, rather than an explicit constraint function $\widetilde{m}(n)$.

The knowledge of $\widetilde{m}(n)$, which includes information about the size and shape of the support $S$, constitute the direct-space constraint in our reconstructions.

\subsection{Noisiness of constraints.}

To prepare for systematic studies of reconstruction feasibility, we classify our diffractive imaging simulations using convenient signal-to-noise parameters. One such consideration is the photon-shot-noise in diffraction data.

Photon-shot-noise is related to the average number of scattered photons per pixel $\muT$, regardless of whether it came from the charge or magnetization distribution (refer to \eeqref{eqn:ScatteredIntensity}). Increasing $\muT$ ought to improve the chances of reconstructing the total scattering distributions. However, using $\muT$ as a signal-to-noise parameter is too optimistic since we are only interested in recovering the magnetization distribution \footnote{The size and shape of the beamstop also affect $\muT$ without practical significance to reconstruction success.}. Consequently, one must still isolate the magnetization distribution from the total scattering distributions, even if the latter is correctly determined (i.e. to extract the magnetization Fig. \ref{fig:MagAndCharge}a from only Fig. \ref{fig:MagAndCharge}c).

To appropriately characterize the noisiness of the photon data to our goal, instead of the total scattered power $\muT$, we use $\mu_m$: the average number of photons scattered due to the magnetization in each pixel within the support $S$. In experiments, $\mu_m$ can be estimated directly from magnetic elastic scattering amplitude $f^M$, photon flux and exposure time of diffraction measurement. In our simulations, $\mu_m$ is computed as 
\begin{equation}
\mu_m = \frac{\phi}{N_{S}} \, \sum_{{\bf q}} \left| m({\bf q}) \right|^2 \; ,
\end{equation}
where $m({\bf q})$ is the discrete Fourier transform of the magnetization distribution $\mr$; $N_{S}$ is the number of support pixels; $\phi$ is the same scalar in \eeqref{eqn:ScatteredIntensity} which we vary to give the desired total number of scattered photons; $f^M$ is again set to unity inconsequentially. The product $\mu_m N_S$ corresponds to the total number of photons scattered per pulse in the absence of charge scattering. 

The other noise consideration comes from charge scattering. We assume that the specimen's random charge distribution is unknown, which results in a harder reconstruction problem. As a consequence, the model magnetizations in Fig. \ref{fig:BinaryContrast} will not agree with those in the total scattering amplitudes of \eeqref{eqn:SimpleCrossSection}, which includes the charge distribution. Essentially, this makes our direct-space constraint noisy \footnote{One could include the expected statistics on the charge distribution in Fig. \ref{fig:BinaryContrast}. This will certainly make the direct-space and Fourier constraints more compatible, potentially improving the reconstruction success rate. Even having included the charge statistics it may still be fairly challenging afterwards to isolate the magnetization distribution from these reconstructions chiefly because the exact charge distribution is unknown. Smoothing operations can remove charge contrast only if it is small compared to the magnetic contrast.}. Experimental measurement of the charge distribution would certainly reduce this noise and simplify the reconstruction.

\section{Reconstruction algorithm.}

\subsection{Modifying the difference map.}
Seeking the true magnetization distribution is equivalent to finding the intersection of the Fourier and direct-space constraint sets. Such intersections, or {\it solutions}, can be discovered using an iterative constraint-satisfaction algorithm: the difference map \cite{Elser:2007}, which uses simple projections to these two constraints ($P_D$, projection to direct-space constraint; $P_F$, projection to Fourier constraint).

The difference map algorithm accelerates the discovery of a solution, primarily by reducing the dimension of the search space \cite{Elser:2007}. It is also particularly efficient in extricating the iterate from near intersections (false solutions) to prevent the search from stalling. However, the difference map algorithm was optimized for noiseless constraints sets with true intersections \cite{Elser:2003}. 

Unfortunately, photon and charge scattering noise distorts our measurement of the true Fourier constraint, demoting its intersections with the direct-space constraint to near intersections, from which iterates are jettisoned. This prohibits the search from reaching the true magnetization distributions encoded in these near intersections. 

To increase its reconstruction success rate, the difference map was modified to improve the stability of the iterate around a near intersection. This is accomplished by an intermediate step to the iteration $m_n \to m_n^{\prime} \to m_{n+1}$ (where the iteration number $n=0,1,2,\ldots$), which keeps the iterate close to the Fourier constraint \footnote{We prefer the iterate to orbit near the Fourier constraint since it is a direct experimental measurement of a particular magnetization distribution, as opposed to the direct-space constraint which is a broader description of the ensemble of distributions.}: 
\begin{eqnarray}
m^{\prime}_n &=& \alpha \,m_n + (1- \alpha)P_F (m_n) \; ,\nonumber \\
\varepsilon_n &=& P_F \left( \, 2 P_D(m^{\prime}_n) - m^{\prime}_n \, \right)  - P_D( m^{\prime}_n) \; , \nonumber  \\
m_{n+1} &=& m^{\prime}_n + \varepsilon_n \; \label{eqn:DifMap}, 
\end{eqnarray}
with $P_D$ and $P_F$ as the direct-space and Fourier constraint projections respectively and $\alpha$ as the map's modification parameter. The update on the iterate is denoted $\varepsilon_n$, so that it may be referenced concisely in later paragraphs. 

\begin{figure}[t!!]
\centering
\includegraphics[width=3.3in]{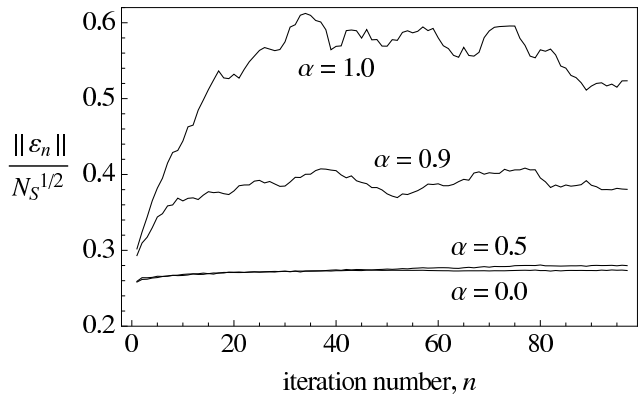}
\caption{Stability of modified difference map, \eeqref{eqn:DifMap}, around a solution. We estimate this stability by tracking how quickly the iterate leaves the solution because of the noisy constraints. Starting from the solution magnetization distribution, the normalized magnitude of the iterate's updates ($||\varepsilon_n||/ N_S^{1/2}$) remains low for $\alpha < 0.5$, indicating iterate stability around a noisy solution. } \label{fig:AlphaStudy}
\end{figure}

In our reconstructions we chose $\alpha = 0$, which substantially improves the iterate's stability (Fig. \ref{fig:AlphaStudy}) while reducing the number of computations in the first step of the algorithm. With $\alpha =1$, \eeqref{eqn:DifMap} reduces to an instance of the original difference map. Appendix \ref{appendix:RAAR} discusses how \eeqref{eqn:DifMap} is similar to the RAAR algorithm in \cite{Luke:2005}. 

\begin{figure}[t!!]
\centering
\includegraphics[width=3.3in]{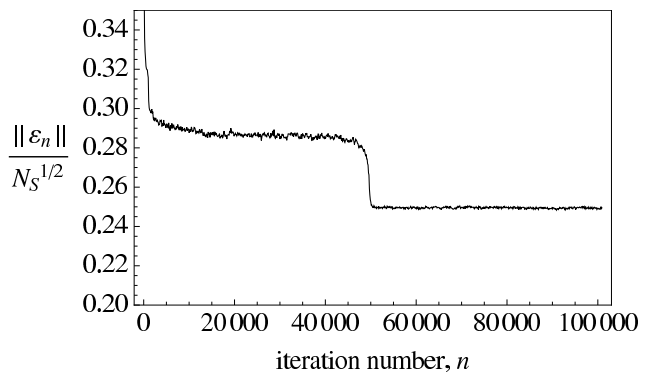}
\caption{Signature of a successful reconstruction. The normalized error metric ($||\varepsilon_n||/ N_S^{1/2}$) suffers a noticeable transition at iteration number $n \approx 50000$ during the successful reconstruction of the distribution in Fig. \ref{fig:LargeRecon}b. }\label{fig:ErrorPlot}
\end{figure}

The modified difference map is iteratively applied to a random, initial magnetization distribution $m_0$. The norm of the map's update $||\varepsilon_n||$, which we term the {\it error metric}, measures the average change of the iterate during the search. When the error metric drastically declines, it indicates that the difference map updates have experienced a dynamic transition and the search has likely converged (Fig. \ref{fig:ErrorPlot}). Because of the inherent noise in the constraints, the error metric will never vanish as it would, had an intersection of the two constraints been found in the noiseless case. When a noticeable transition in $||\varepsilon_n||$ occurs and is stable, we harvest the {\it candidate solution} of the magnetization distribution, $P_D\left( P_F \left( \, 2 P_D(m^{\prime}_n) - m^{\prime}_n \, \right) \right) $. The correctness of this candidate solution is tested when compared against other candidate solutions from different, random, initial iterates $m_0$. Consistent recovery of nearly identical candidate solutions, up to an overall multiplicative sign or spatial inversion, from random restarts asserts their credibility as the true magnetization distribution. One can smooth out the fluctuations between the candidate solutions by averaging them. 

In searches using the noisiest photon data, the error metric $||\varepsilon_n||$ will never show a clear transition. In such cases, recovering the true magnetization distribution is plainly impossible. Nonetheless, we can still evaluate the search results, however wrong they may be. From \eeqref{eqn:DifMap}, notice that $||\varepsilon_n||$ also measures the distance between two points on the two constraints: $P_F \left( \, 2 P_D(m^{\prime}_n) - m^{\prime}_n \, \right)$ and $P_D( m^{\prime}_n)$. Hence the minimum $||\varepsilon_n||$ during a search signals the nearest distance between the two constraints --- the best alternative to discovering an intersection. Unlike more robust candidate solutions with less noisy data, these {\it faux solutions} are never repeated with random restarts. 

\subsection{Projection to direct-space constraint.}\label{ssec:DSProj}

The projection to the direct-space constraint, $P_{R}(m)$, comprises the following operations on $m$:
\begin{enumerate}
\item{set all values of $\mr$, for ${\bf r}$ outside $S$, equal to zero;}  
\item{$m(n)$ replaced by the magnetizations $\widetilde{m}(n)$ shifted and scaled to have the same mean and variance as $m(n)$ before the projection.} \label{directProj:MeanVar} 
\end{enumerate}
Step \ref{directProj:MeanVar} allows the mean scattering amplitude $\langle \fCr\rangle$ and the magnetic contrast $\Delta_m$ to be indirectly constrained by the diffraction data  \footnote{The mean charge scattering amplitude $\langle \fCr\rangle$ is non-critical to the reconstruction since it constitutes mainly the missing intensities in the data where the diffraction intensities from the sample's magnetization is low (see Fig. \ref{fig:Data}).}. 

In actual experiments where the magnetization constraint function $\widetilde{m}(n)$ is not readily available or simulated, one could instead project $m(n)$ to a class of parametrized magnetization functions, where the projection determines the best parameter. When even this is impossible, imposing only the key features of $\widetilde{m}(n)$ on $m(n)$ may be sufficient (Appendix \ref{appendix:SimpleDirectSpaceConstraint}).

\begin{figure}[t!!]
\centering
\includegraphics[width=1.1in]{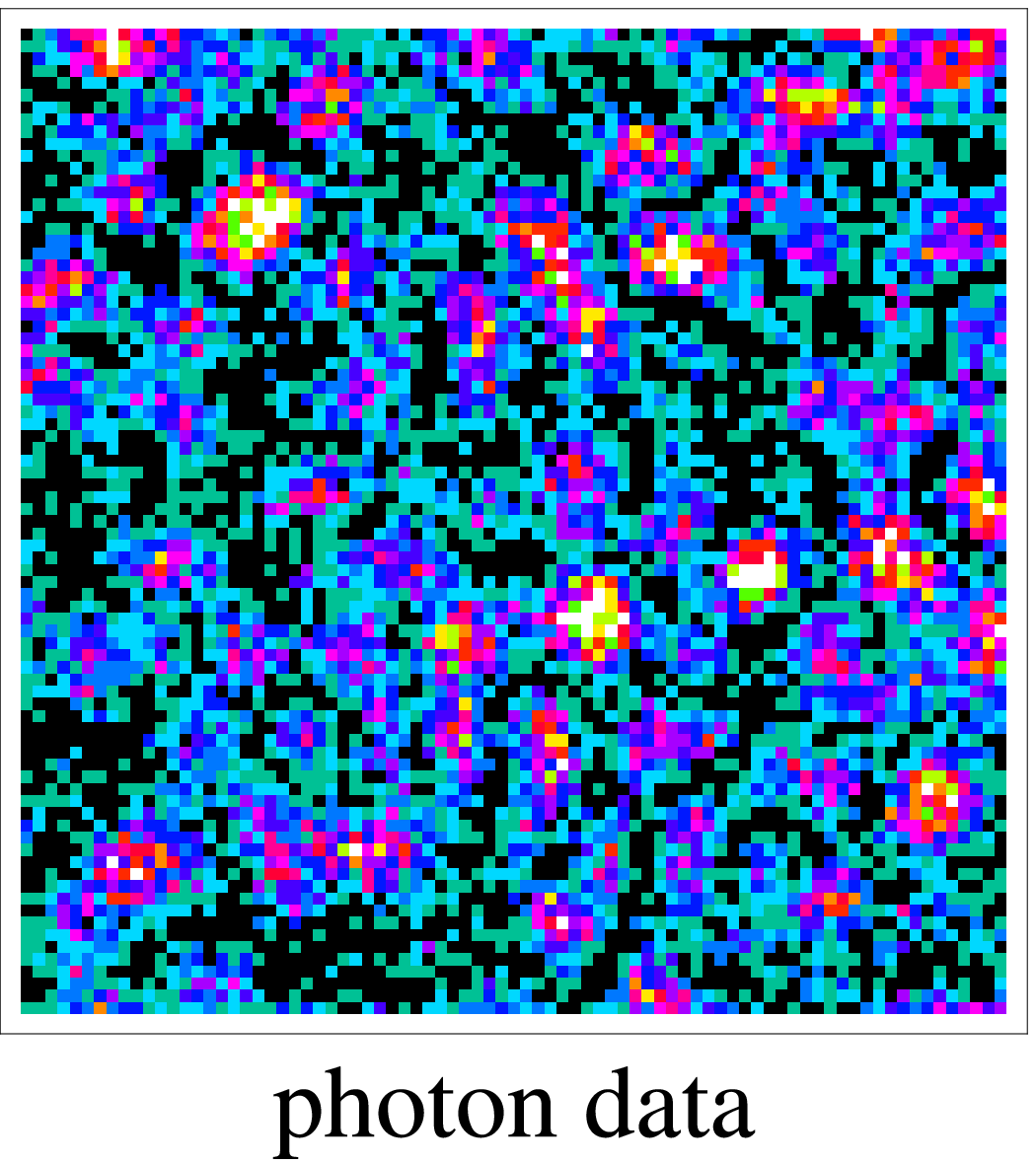}
\includegraphics[width=1.1in]{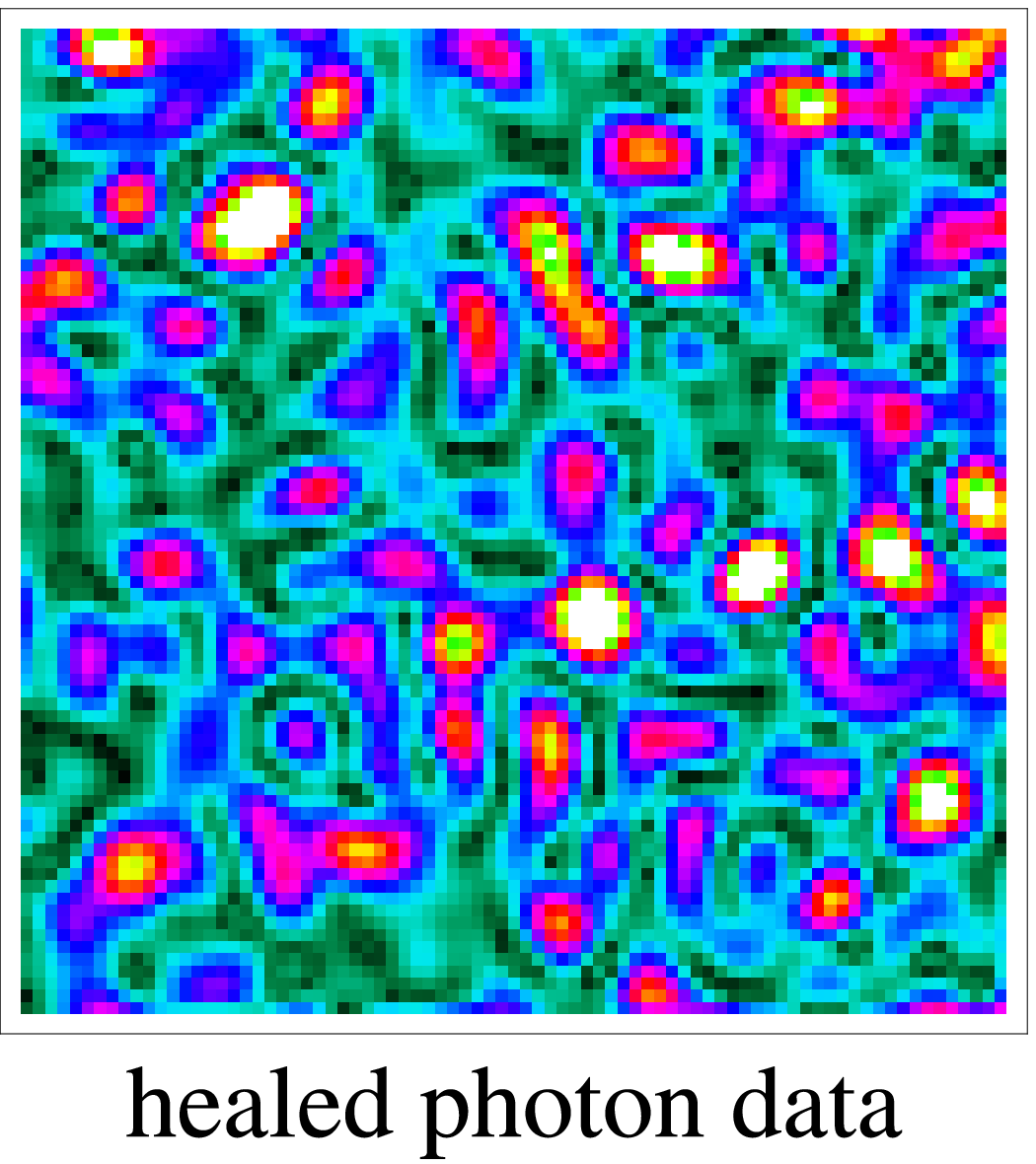}
\includegraphics[width=1.1in]{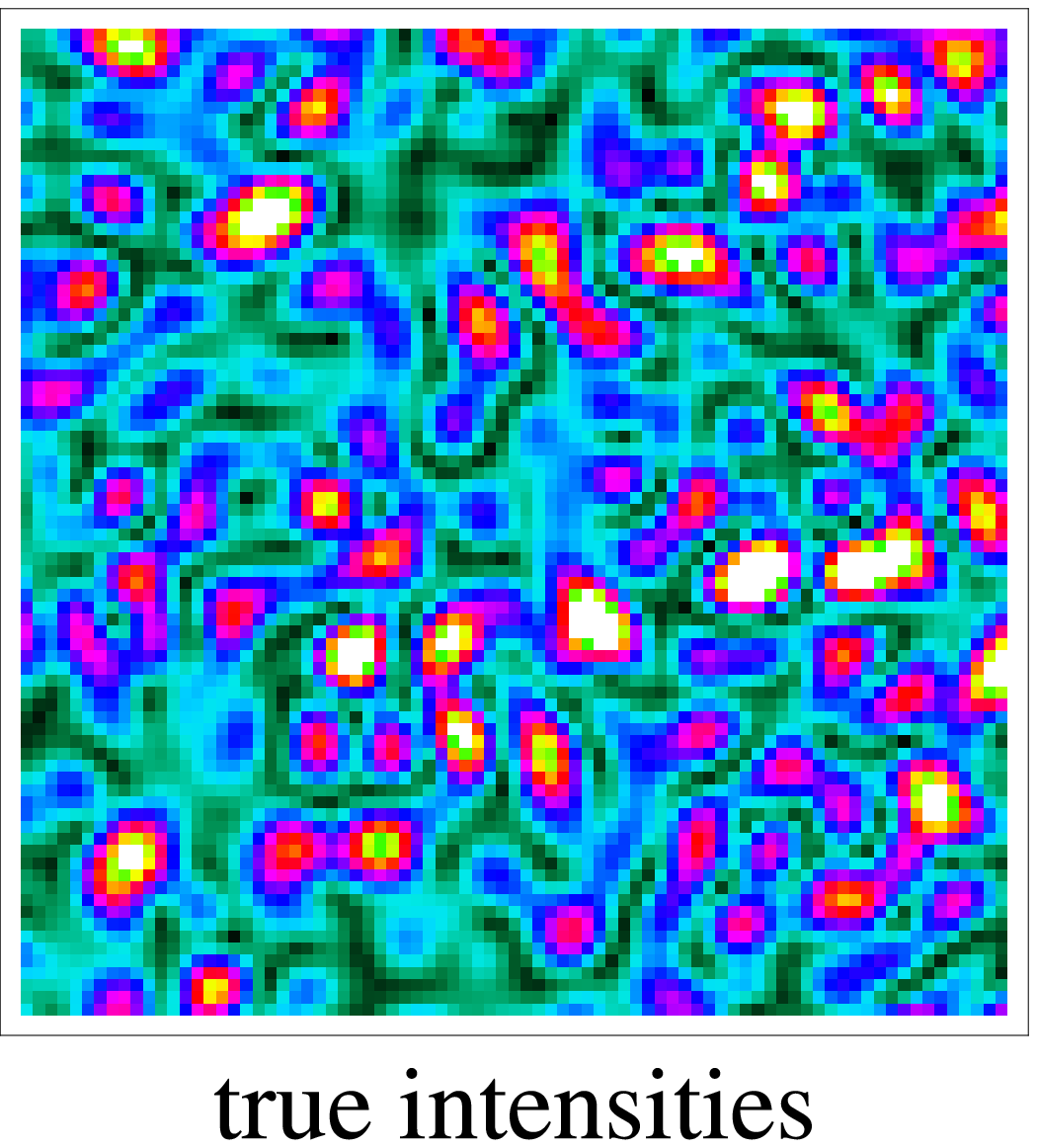}
\caption{(color online). Speckle-healing by applying an autocorrelation support constraint to the photon data during reconstruction. We show this healing of a magnified section of the photon data in Fig. \ref{fig:Data}. }\label{fig:SpeckleHealing}
\end{figure}

\subsection{Projection to Fourier constraint.}

Before discussing the projection to the Fourier constraint, $P_{F}(m)$, we describe a modification to the diffraction data which lowers the photon-shot-noise using the direct-space constraint. If the scattering distribution is contained within a direct-space support $S$, the Fourier transform of the diffraction intensities --- or the autocorrelation of the direct-space scattering distribution --- should be contained within the autocorrelation support $S_A$. 

We can lower the noise in the diffraction data using our knowledge of the support, hence constraining the photon data to have the expected {\it speckles}. We did so by applying an autocorrelation support constraint to the Fourier transform of the photon data --- setting all values outside $S_A$ in the data's Fourier transform to zero. Empirically, this {\it speckle-healing operation} increases the R-factor between the processed photon data and the true intensities (see Fig. \ref{fig:SpeckleHealing}). 

However, the missing data within the beamstop may confuse speckle-healing. These central Fourier amplitudes are indirectly constrained by the diffraction data and after numerous iterations the difference map iterate proposes preliminary intensities for them. We replace the missing photon data with these preliminary intensities before applying the speckle-healing operation. In our reconstructions, the photon data was healed this way every 1000 iterations, which were then used to constrain iterations until the next healing. When the reconstruction converges under this adiabatic healing process, we are assured that it is still compatible with the photon data. 

With this adiabatic speckle-healing procedure in effect, the projection to the Fourier constraint, $P_{F}(m)$, comprises the following operations on the iterate's Fourier transform $m({\bf q})$:
\begin{enumerate}
\item{set the amplitudes of $m({\bf q})$ outside the beamstop to the square root of the speckle-healed photon data, while retaining the phases of $m({\bf q})$; }
\item{the $m({\bf q})$ values within the beamstop are unchanged.}
\end{enumerate}

\section{Feasibility.}

\begin{figure}[t!]
\centering
\includegraphics[width=3.5in]{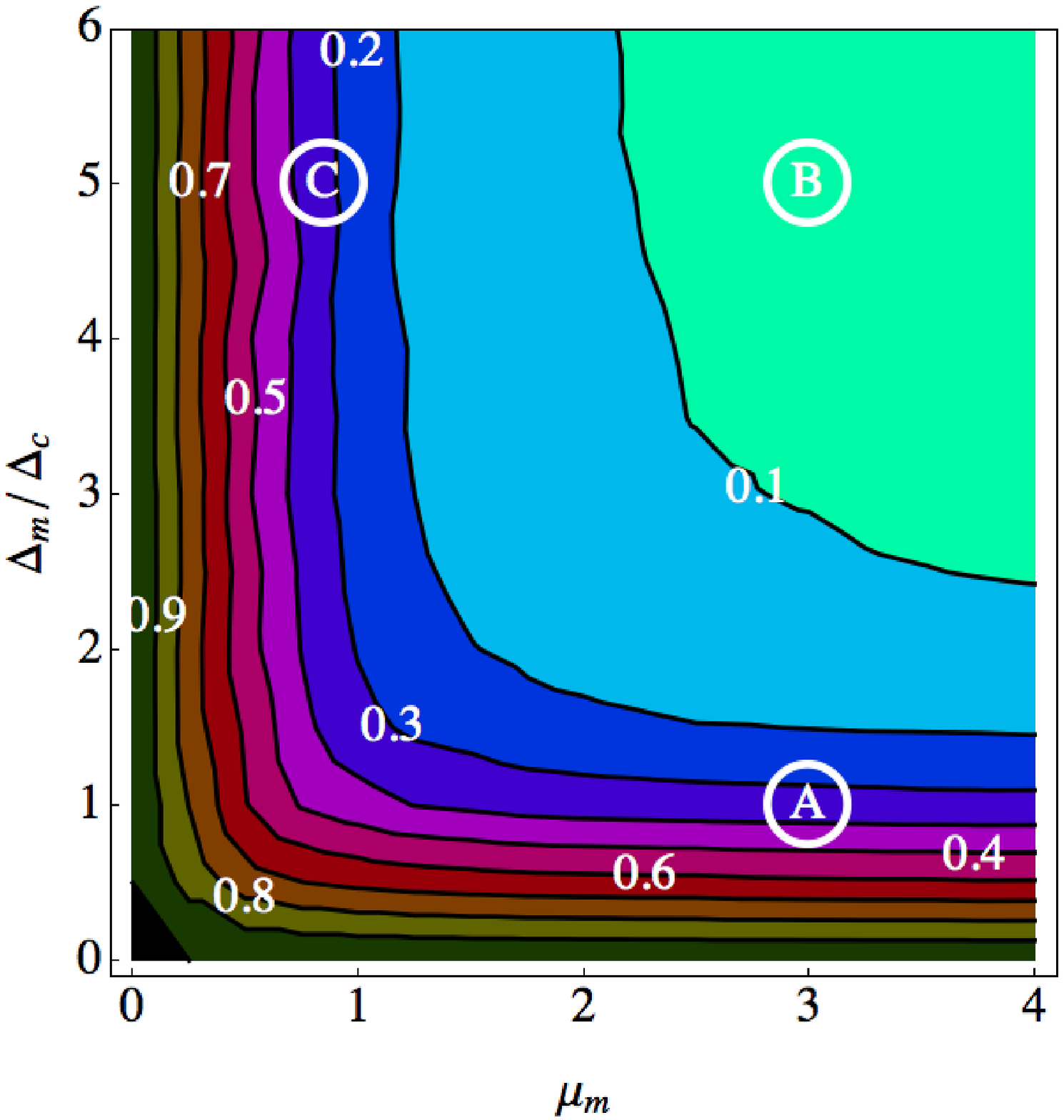}
\includegraphics[width=1.6in]{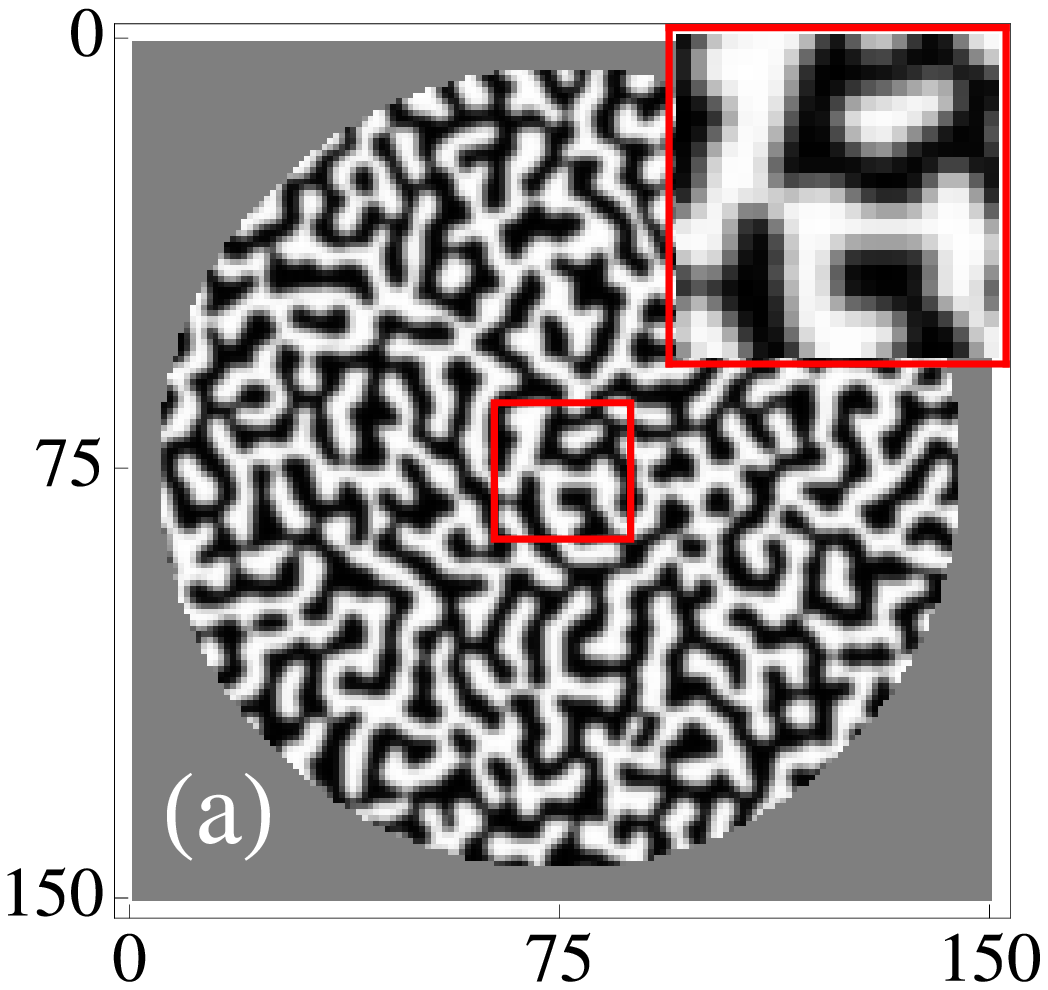}
\includegraphics[width=1.6in]{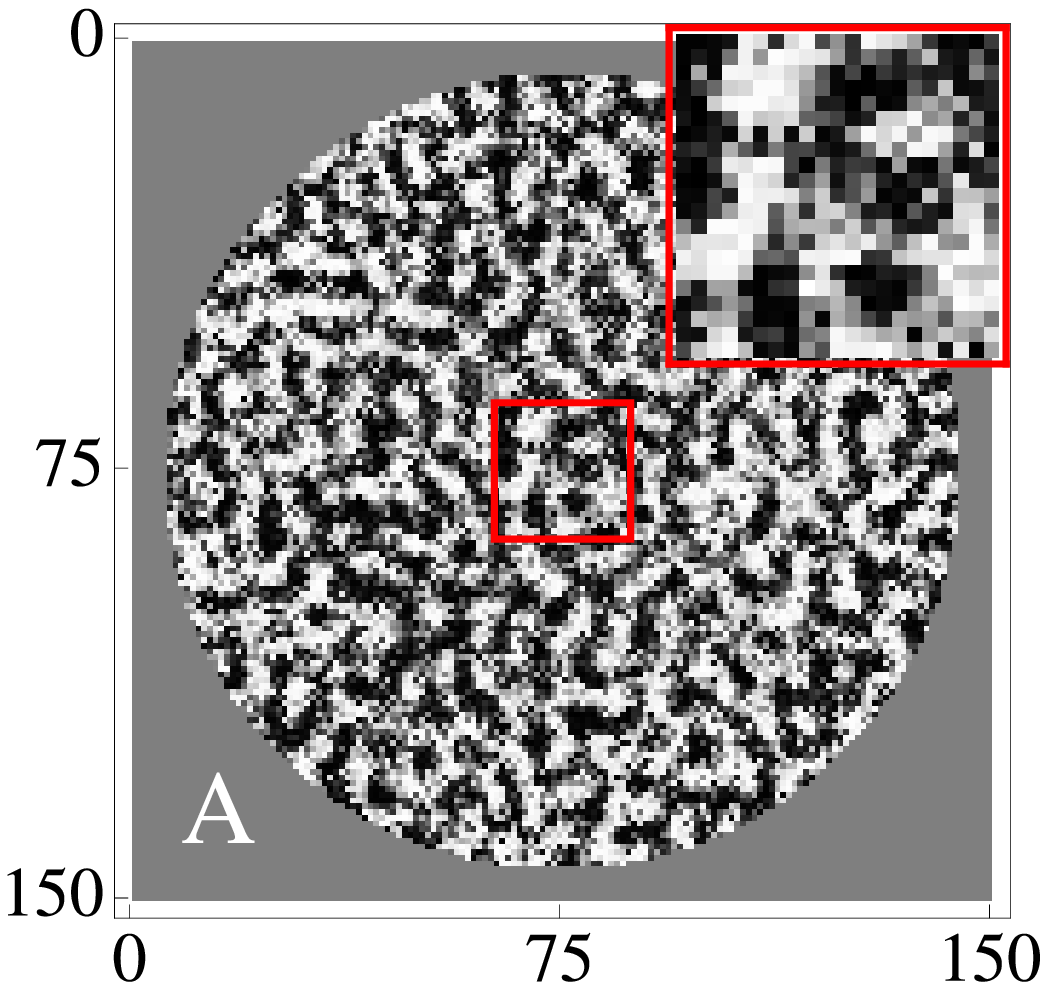}
\includegraphics[width=1.6in]{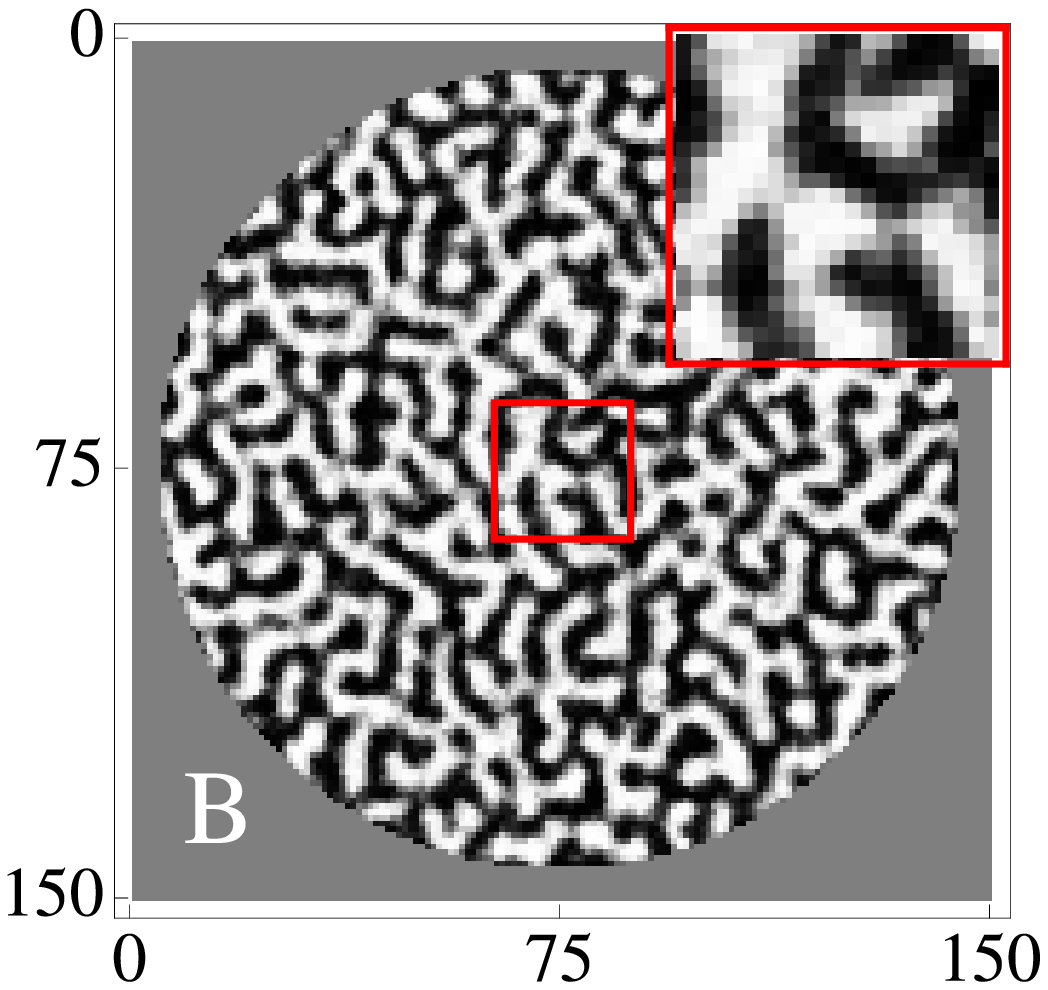}
\includegraphics[width=1.6in]{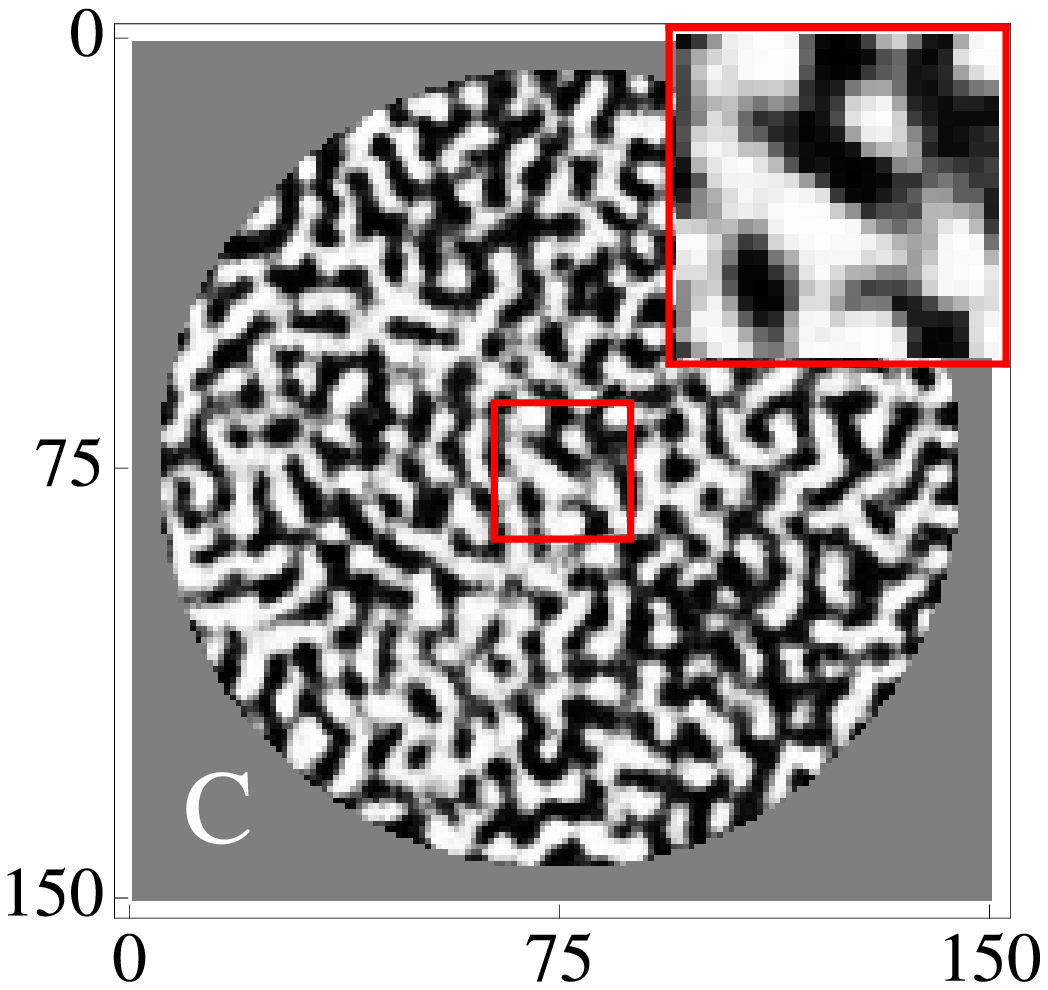}
\caption{(color online). Feasibility of diffractive imaging at various noise levels. The {\bf top} panel shows a contour plot of the reconstruction deviation $\delta$ (in \eeqref{eqn:Deviation}) generated from many simulated reconstructions of the pure magnetization distribution similar to {\bf (a)} as the signal-to-noise parameters ($\Delta_m/\Delta_c$ and $\mu_m$) were independently varied. Panels {\bf A, B} and {\bf C} show reconstructions of (a) subject to corresponding signal-to-noise parameters marked in the top panel.}\label{fig:ContourPlot}
\end{figure}

\begin{figure}[t!]
\centering
\includegraphics[width=3.3in]{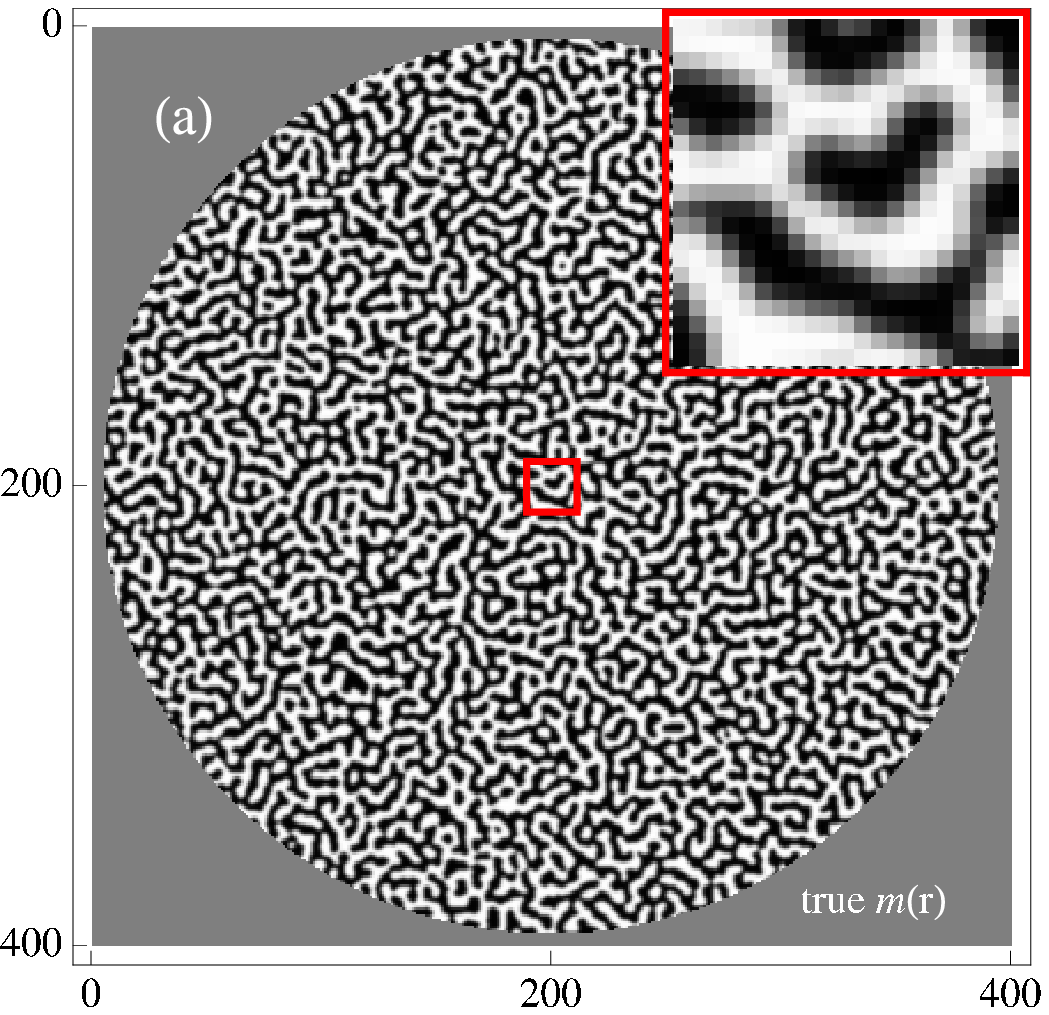}
\includegraphics[width=3.3in]{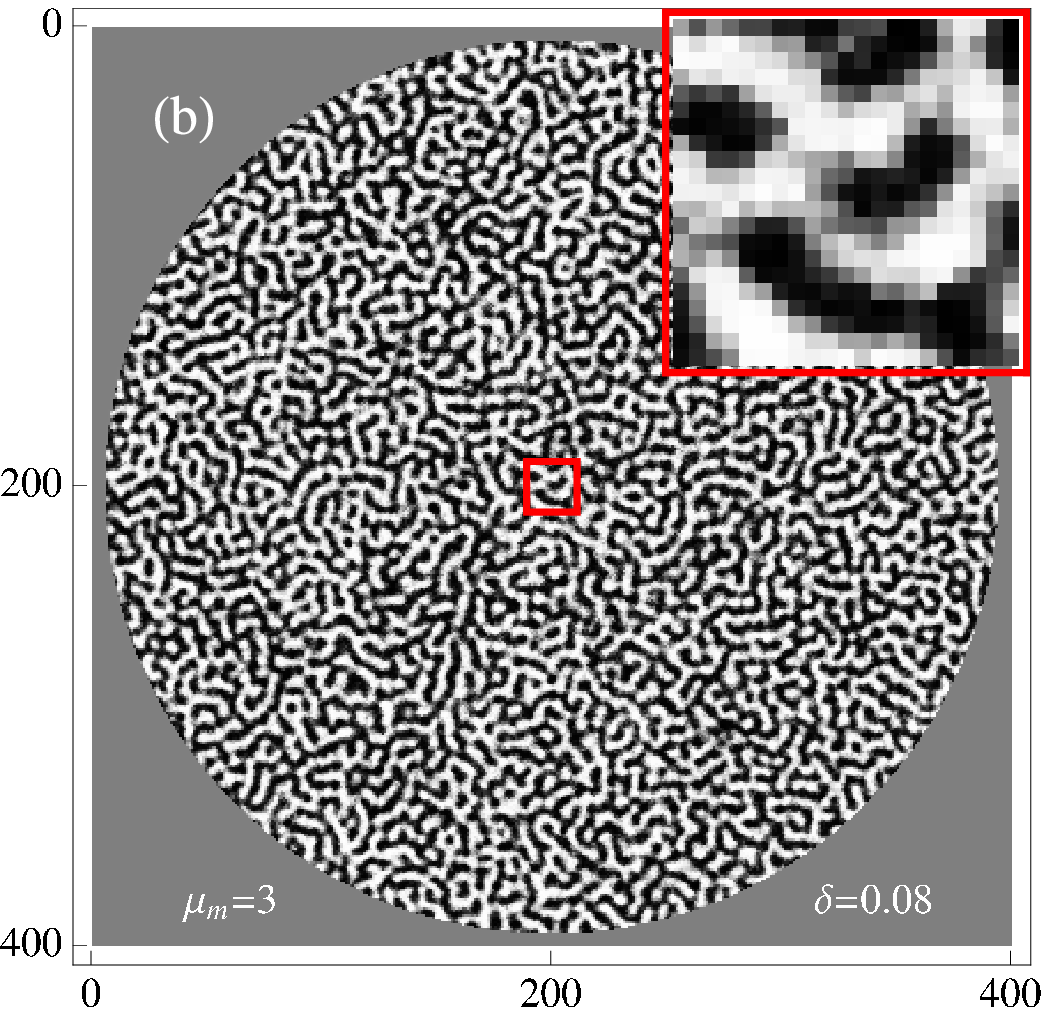}
\caption{In panel {\bf (b)}, a modified difference map reconstruction of a simulated magnetization distribution, panel {\bf (a)}, with signal-to-noise parameters corresponding to point B in Fig. \ref{fig:ContourPlot}. The diffraction data used for this reconstruction is shown in Fig. \ref{fig:Data}. If the magnetic domains are 170 nm wide, then having only three photons scattered by the magnetic contrast within each 34 nm pixel was sufficient to reconstruct the pattern in the bottom panel.}\label{fig:LargeRecon}
\end{figure}

\begin{figure}[t!]
\centering
\includegraphics[width=1.65in]{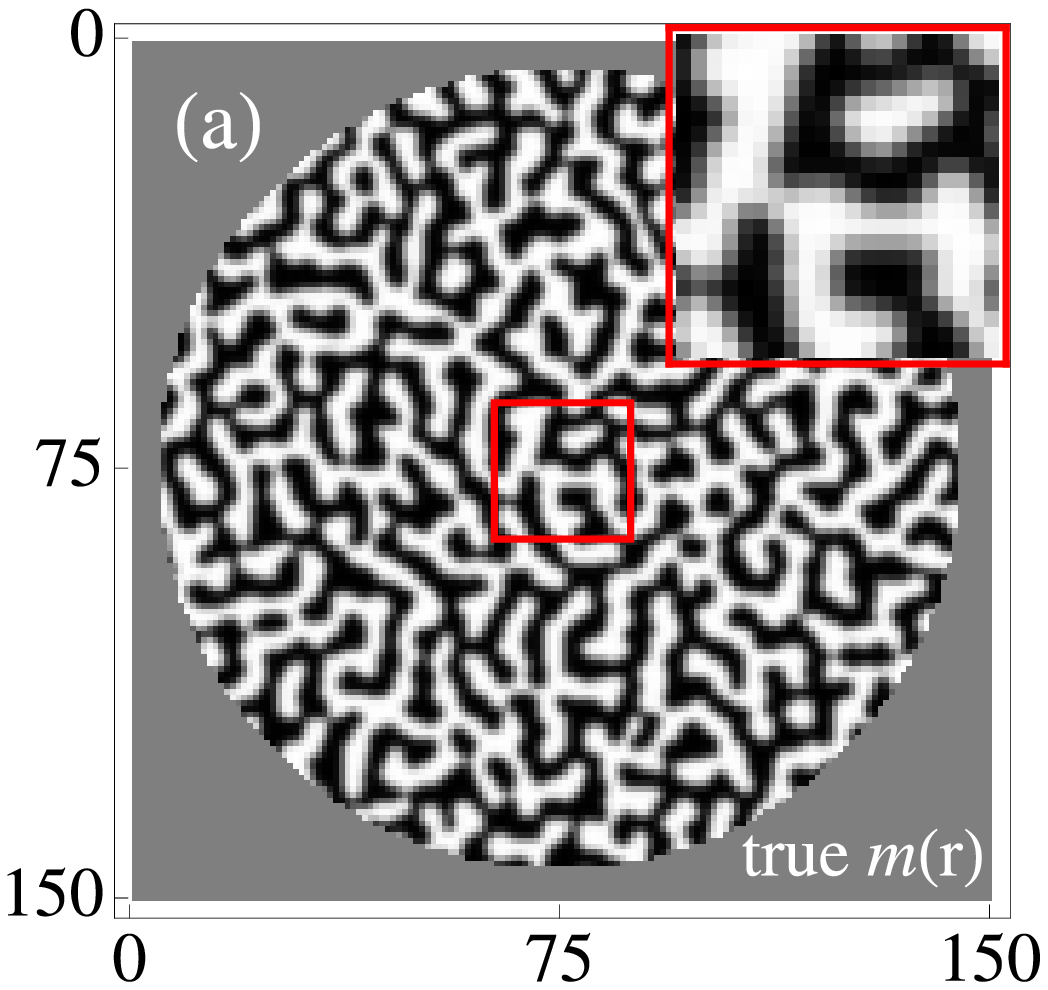}
\includegraphics[width=1.65in]{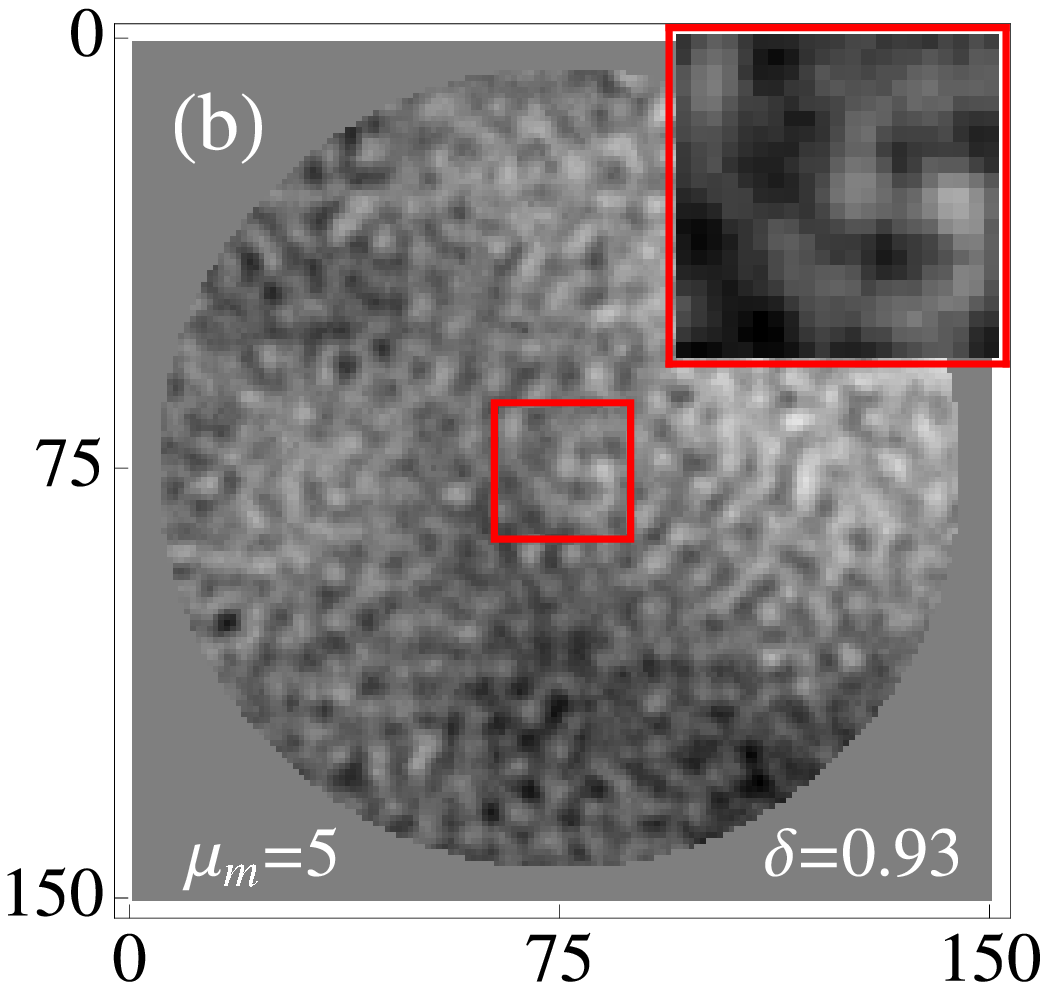}
\includegraphics[width=1.65in]{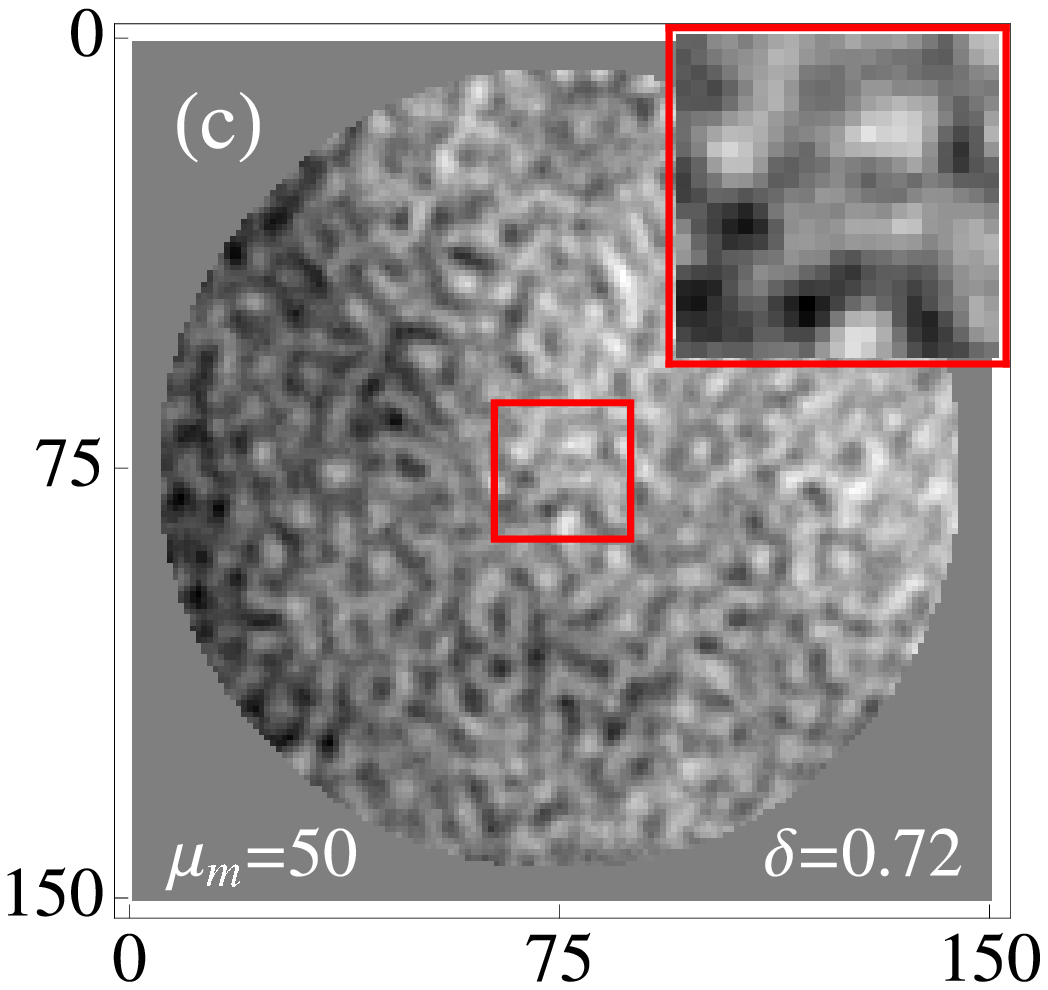}
\includegraphics[width=1.65in]{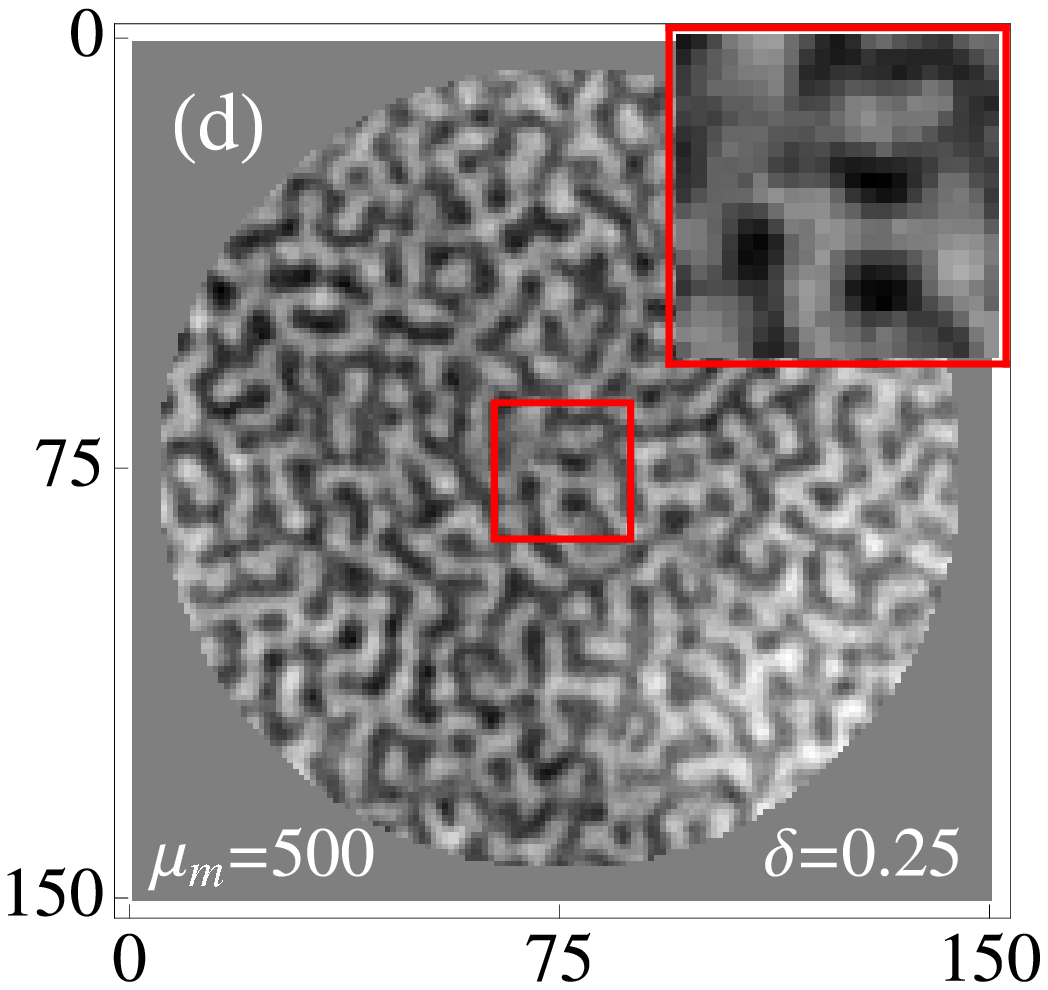}
\includegraphics[width=3.3in]{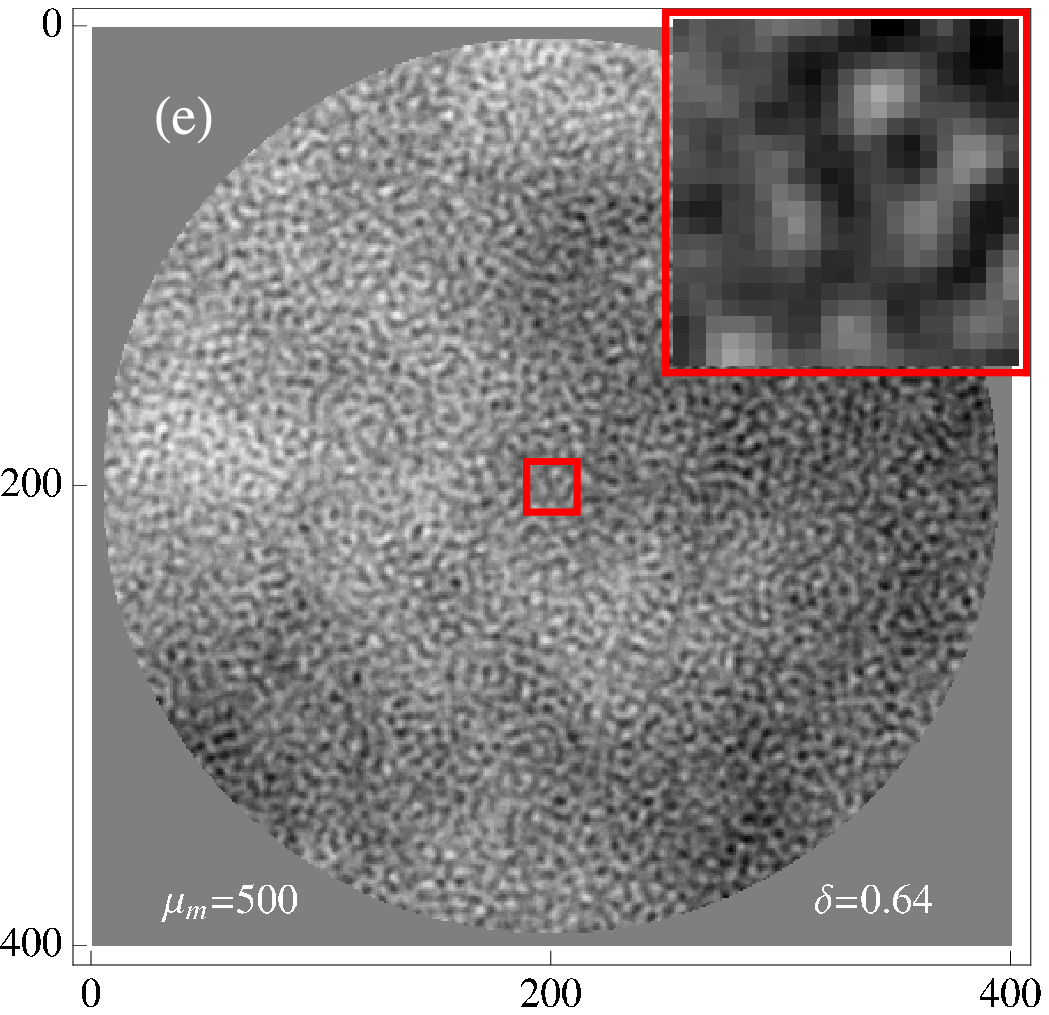}
\caption{Simulated reconstructions with Fourier transform holography (FTH). The panels {\bf (b), (c), (d)} show how FTH require more scattered photons for acceptably low reconstruction deviations. Panel {\bf (e)} is a FTH reconstruction of Fig. \ref{fig:LargeRecon}a. Panels (d) and (e) together show that reconstructions worsen with increasing support size. The relative magnetic contrast was $\Delta_m/\Delta_c = 10$ in these reconstructions.}\label{fig:FTH}
\end{figure}

\subsection{Difference map reconstructions.}

Unlike an actual experiment, the true magnetization distributions are known in our simulated experiments. This allows us to directly compare the reconstructions $m^{\text{rec}}$ with the true distribution $m^{\text{true}}$ within the support $S$ via the following deviation measure:
\begin{eqnarray}
\delta &=& \frac{1}{2}\, \sum_{{\bf r} \in S} \left( u^{\text{rec}}({\bf r}) - u^{\text{true}}({\bf r}) \right)^2 \label{eqn:Deviation}\\ 
u({\bf r}) &=& \frac{m({\bf r})}{\left( \sum_{{\bf r} \in S} m({\bf r})^2 \right)^{1/2}} \; .
\end{eqnarray}
The deviation $\delta$ is proportional to the square of the distance between $u^{\text{rec}}$  and $u^{\text{true}}$, which are the respective distributions normalized as unit vectors. Allowing for an overall sign in the reconstructed magnetization, deviations lie within the range $0 \leq \delta \leq 1$.

In our simulations, reconstructions with $\delta < 0.2$ were consistently recovered from random restarts. In actual experiments, only the consistency test is available to evaluate the reliability of the reconstructions. We deem such repeatable reconstructions to be {\it successful}.

We systematically studied the performance of our reconstruction algorithm when we varied the two signal-to-noise parameters: relative magnetic contrast $\Delta_m / \Delta_c$ and the average number of photons scattered from the magnetic distribution $\mu_m$. As Fig. \ref{fig:ContourPlot} indicates, increasing $\Delta_m / \Delta_c$ and $\mu_m$ improves the accuracy of the reconstructions. However the same figure shows that the effects of suppressing either $\Delta_m / \Delta_c$ or $\mu_m$ are qualitatively different --- lowering one variety of noise can not compensate for the reconstruction errors caused by the other.

Magnetization distributions shown in Fig. \ref{fig:LargeRecon} are routinely reconstructed with low deviations $\delta \leq 0.1$ given remarkably noisy data typical of Fig. \ref{fig:Data}: relative magnetic contrast $\Delta_m/\Delta_c = 5$ and average scattered photons due to the magnetization in each pixel $\mu_m = 3$. The deviation of reconstructions from the true domain pattern at various noise levels is numerically computed in Fig. \ref{fig:ContourPlot} and appears to be independent of the support size at a constant domain resolution (the reconstructions in Fig. \ref{fig:ContourPlot}B and Fig. \ref{fig:LargeRecon}b suffered comparable noise levels). 

Reconstructions with the unmodified difference map, $\alpha = 1$ in \eeqref{eqn:DifMap}, do not converge within the range of noise parameters in Fig. \ref{fig:ContourPlot}: neither in the sense of achieving a dynamic transition in the error metric (Fig. \ref{fig:ErrorPlot}) nor repeatability given random restarts. We witness this lesser performance even with reconstructions using the modified difference map when we omit either the sorted-value magnetization constraint in the direct-space projection or the speckle-healing procedure or both.

\subsection{Comparison with Fourier transform holography.}

To provide perspective, we compared our reconstructions with those from simulated Fourier transform holography (FTH) in Fig. \ref{fig:FTH}. In FTH, the domain pattern is obtained directly from its cross-correlation with an aptly machined reference pinhole \cite{Eisebitt:2004}. This cross-correlation is obtained from a simple Fourier transform of the measured diffraction intensities without the need for phase retrieval.

To make the comparison more compelling, we provided our FTH simulations with the following advantages over the non-holographic method: 
\begin{enumerate}
\item{} noisy diffraction signal within the beamstop region was provided;
\item{} single-pixel reference pinhole for highest possible reconstruction resolution (pinhole diameter roughly 34 nm if magnetic domains are 170 nm wide) whereas the pinhole in \cite{Eisebitt:2004} which had an effective X-ray transmission diameter of approximately 100 nm.
\end{enumerate} In our simulated FTH reconstructions, the ratio of the number of photons scattered by the magnetic contrast in each support pixel to the number which pass through each pixel of the reference pinhole is 1:50, as estimated from \cite{Eisebitt:2004}. 

At the low signal-to-noise levels of Fig. \ref{fig:ContourPlot}, the low deviation reconstructions using our proposed non-holographic diffractive imaging technique are out of the reach of our implementation of FTH (compare Figs. \ref{fig:ContourPlot} to \ref{fig:FTH}). Fig. \ref{fig:FTH} also illustrates that our FTH reconstructions worsen with increasing support size because the photon fluence through the pinhole does not increase commensurately \footnote{One could average the cross-correlations of the magnetic contrast with multiple references to improve the FTH reconstructions as demonstrated in reference \cite{Schlotter:2006}. The signal-to-noise ratio of these averaged FTH reconstructions is expected to increase with the square root of the number of references. In our trials, the deviation of the FTH reconstruction Fig. \ref{fig:FTH}e falls to 0.15 when the number of references is increased from 1 to 16. Although this deviation is acceptably low, for the same performance it still requires roughly 150 times more photons than our non-holographic technique.}. This reflects the typical situation in microscopy that higher resolution necessitates a smaller field of view. Non-holographic diffractive imaging does not suffer this size dependency since only the noise per support pixel is important. While non-holographic diffractive imaging does not need the experimental fabrication of a small reference object and can use the beam's spatial coherence more efficiently via a tighter X-ray focus, it requires accurate knowledge of the support \cite{Thibault:Thesis}. These differences between the techniques make the non-holographic phase retrieval approach demonstrated here of particular interest for situations where the signal is too noisy for successful FTH of extended magnetization distributions.

\section{Conclusions.}
Ultrafast imaging of magnetic nanostructures is presumably possible within the noise limits predicted by Fig. \ref{fig:ContourPlot}. This, of course, is valid only in the absence of other varieties of noise. Our study is limited to magnetic imaging without prior measurement of the random charge distribution. We speculate that the reconstruction noise limits would improve if the specimen's charge distribution, which may fluctuate, were available. 

Certainly, imposing ensemble properties of the domain patterns in our reconstruction algorithm allows magnetic imaging with remarkably noisy data. Although our reconstructions use the sorted-value magnetization constraint, an approximate knowledge of this constraint may be satisfactory (see Appendix \ref{appendix:SimpleDirectSpaceConstraint}). Despite restricting our simulations to a small ensemble of domain patterns, the methods we used to reconstruct these patterns should be valid for imaging a larger ensemble of ferromagnetic contrast that differ only qualitatively from ours.

Our success with the modified difference map, \eeqref{eqn:DifMap}, suggest its relevance to constraint-satisfaction problems that suffer from imprecise or noisy constraints. Similarly, the speckle-healing procedure in this paper is pertinent to recovering missing global information common in diffractive imaging. 

\section{Acknowledgements.}
This work was supported by CHESS through NSF and NIH/NIGMS via NSF award DMR-0225180. We thank Yoav Kallus, for his indispensable insights with regards to the domain-generation prescription, and Victor Lo for pointing out the similarity of our modified difference map to the RAAR algorithm \cite{Luke:2005} (see appendix \ref{appendix:RAAR}). We also thank Jyoti Mohanty for his assistance regarding magnetic domain theory.

\appendix

\section{Similarity of modified difference map to relaxed averaged alternating reflections algorithm.} \label{appendix:RAAR}

The modified difference map in \eeqref{eqn:DifMap} resembles the relaxed averaged alternating reflections algorithm (RAAR) used in iterative phase retrieval \cite{Luke:2005}. Like the modified difference map, RAAR was designed to stabilize iterates in the domain of attraction of a solution given noisy diffraction data. To see their resemblance, we combine the last 2 lines of \eeqref{eqn:DifMap} as a single operation $D$:

\begin{eqnarray}
m_n^{\prime} &=& \alpha \, m_n + (1 - \alpha) P_F(m_n) \; , \\
m_{n+1} &=& D(m_n^{\prime}) \; . 
\end{eqnarray}
The first step in the next iteration would be 
\begin{eqnarray}
m_{n+1}^{\prime} &=& \alpha \, m_{n+1} + (1 - \alpha) P_F(m_{n+1}) \; \\
&=&\alpha \, D(m_{n}^{\prime}) + (1 - \alpha) P_F( D(m_n^{\prime})) \; ,
\end{eqnarray}
which is similar in structure to the RAAR update:
\begin{equation}
m_{n+1} = \alpha \, D(m_n) + (1 - \alpha) P_F(m_n) \; .
\end{equation}

\section{Symmetry-and-boundedness constraint.} \label{appendix:SimpleDirectSpaceConstraint}

For cases when the ensemble's sorted-value magnetization $\widetilde{m}(n)$ is unavailable, it could be replaced with a less restrictive value constraint. One such replacement is the magnetization's expected symmetry in \eeqref{eqn:MagSymmetry}. This occurs in magnetic samples of zero net magnetization in the absence of external fields. In addition to this symmetry constraint, the magnetization values must bounded by  $-1 < m(n)< 1$ given our normalization. Magnetization symmetry-and-boundedness together constitute a weaker direct-space value constraint; it is weaker because it includes magnetization functions besides the true one. 

\begin{figure}[t!!]
\centering
\includegraphics[width=3.3in]{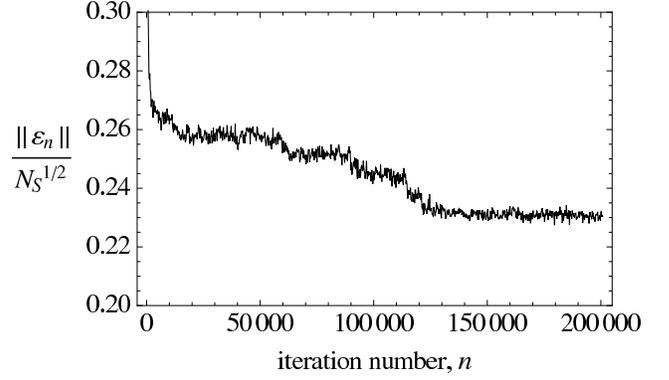}
\caption{Normalized error metric for a successful reconstruction using the weaker symmetry-and-boundedness value constraint. The dynamic transition in the error metric is still visible, though smoothed out because of this less restrictive value constraint.}\label{fig:WeakError}
\end{figure}

\begin{figure}[h!!]
\centering
\includegraphics[width=3.3in]{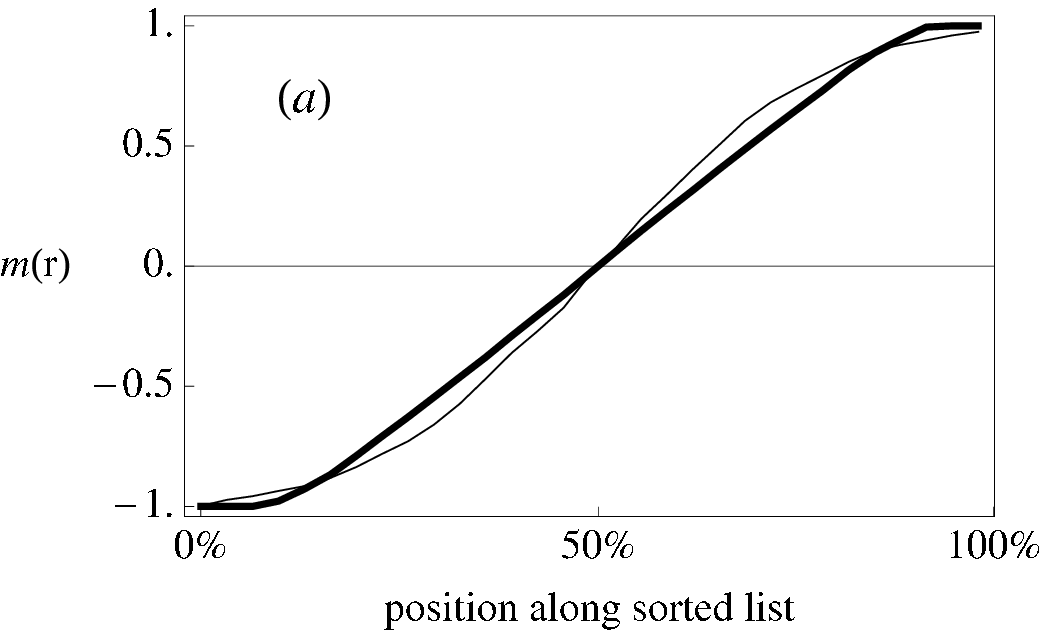}
\includegraphics[width=1.6in]{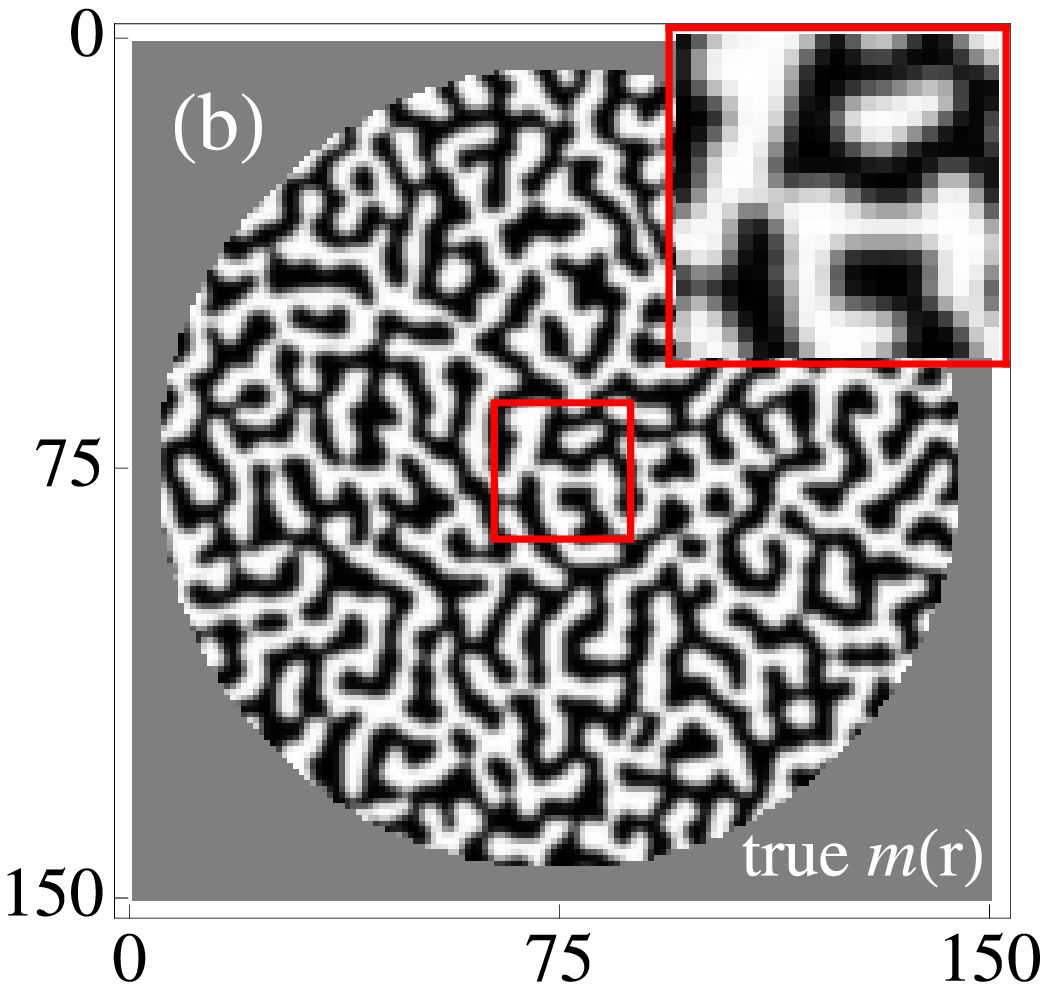}
\includegraphics[width=1.6in]{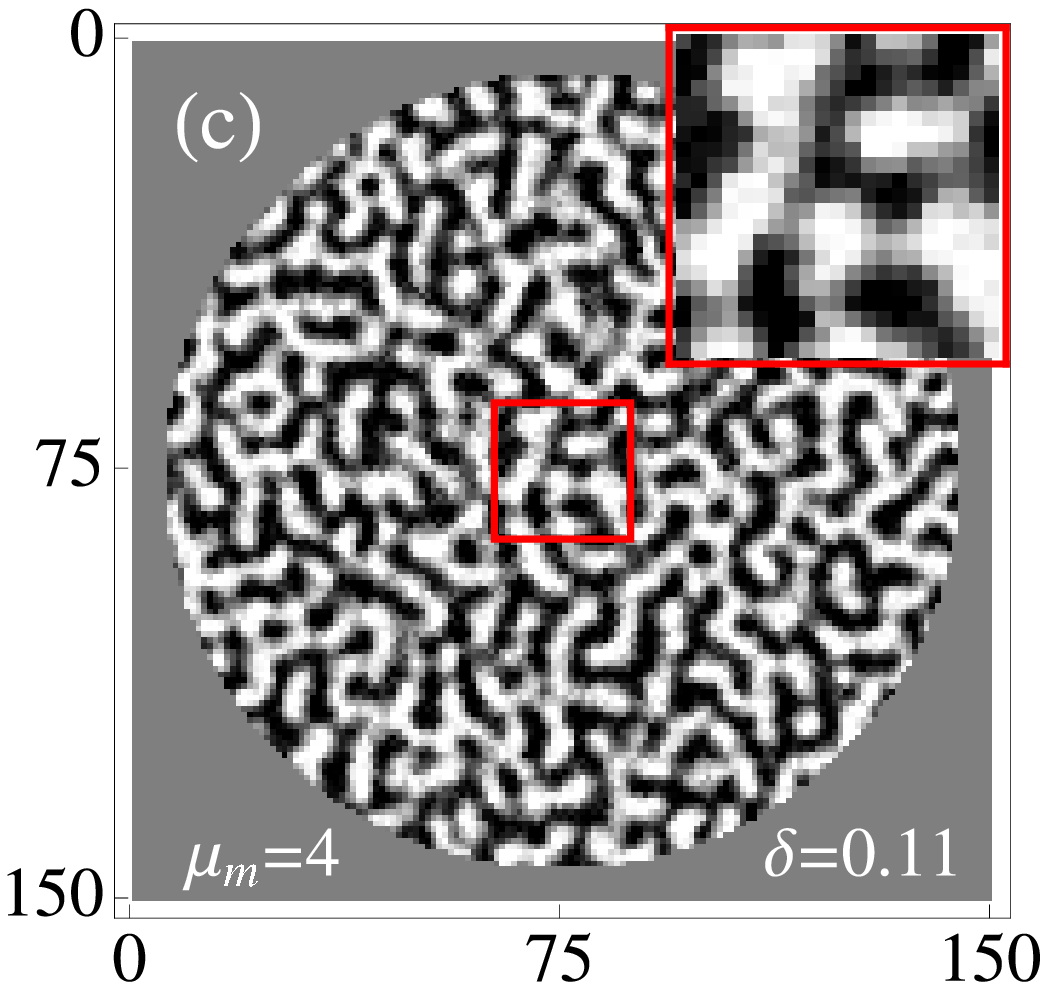}
\caption{A reconstruction using the weaker symmetry-and-boundedness direct-space constraint. In panel {\bf (a)}, the ordered reconstructed magnetizations (dark curve) approximates the true values (light curve). Panel {\bf (c)} shows such a reconstruction of the source domain pattern in panel {\bf (b)}, with $\mu_m=4$ and $\Delta_m/\Delta_c = 6$ (top right corner of contour plot in Fig. \ref{fig:ContourPlot}). Compare panel (c) of this Figure with Fig. \ref{fig:ContourPlot}B, which uses the sorted-value magnetization constraint and reconstructs the pattern with lower deviation while needing slightly fewer photons and tolerating more charge scattering noise.}\label{fig:WeakRecon}
\end{figure}

To test its effectiveness, this weaker constraint was used to reconstruct a domain pattern instead of using its true model magnetization $\widetilde{m}(n)$. This mimics the experimental scenario when one attempts reconstruction even when $\widetilde{m}(n)$ is not completely known beyond its symmetry and boundedness. Projecting to this constraint modifies the direct-space projection of Section \ref{ssec:DSProj}:

\begin{equation}
m(n)\to \left\{
\begin{array}{lc}
\mbox{sgn}(\overline{m}(n)) &\mbox{if }\;  |\overline{m}(n)|>1\\
\overline{m}(n)& \mbox{otherwise,}
\end{array}\right.
\end{equation}
where
\begin{equation}
\overline{m}(n)=\frac{1}{2}\left(m(n)-m(N_S+1-n)\right).
\end{equation}

Practical reconstructions using this weaker direct-space constraint (Fig. \ref{fig:WeakRecon}) require the data to have slightly better signal-to-noise than reconstructions using the ensemble's list of model magnetizations (Fig. \ref{fig:ContourPlot}). This is because the weaker constraint permits model magnetizations different from the true one and is hence a lesser guide during our search for the solution.

A weaker constraint also causes smoothing of the transition in the error metric during a successful reconstruction due to the relaxation of the reconstructed magnetizations towards the true magnetization constraint function (Fig. \ref{fig:WeakError}). With more photons and less charge scattering noise, the magnetizations reconstructed using the symmetry-and-boundedness constraint (Fig. \ref{fig:WeakRecon}a) become closer to the true magnetization function. This suggests that one could obtain accurate magnetization functions from low-noise diffractive imaging experiments to be used as constraints for noisier ultrafast imaging. 

We note that the symmetry-and-boundedness constraint has a crucial difference from the sorted-value magnetization constraint: the former does not explicitly reject charge scattering as a source of noise while the latter does.

\bibliographystyle{unsrt}
\bibliography{papers}

\begin{thebibliography}{10}

\bibitem{Saga:1999}
H.~Saga et~al.
\newblock New recording method combining thermo-magnetic writing and flux
  detection.
\newblock {\em Japanese Journal of Applied Physics}, 38:1839--1840, 1999.

\bibitem{Eisebitt:2004}
S.~Eisebitt et~al.
\newblock Lensless imaging of magnetic nanostructures by x-ray
  spectro-holography.
\newblock {\em Nature}, 432(7019):885--8, Dec 2004.

\bibitem{Pierce:2003}
M.~S. Pierce et~al.
\newblock Quasistatic x-ray speckle metrology of microscopic magnetic
  return-point memory.
\newblock {\em Physical Review Letters}, 90(17):175502, May 2003.

\bibitem{Gutt:2010}
C.~Gutt et~al.
\newblock Single-pulse resonant magnetic scattering using a soft x-ray
  free-electron laser.
\newblock {\em Physical Review B}, 81(10):100401, Mar 2010.

\bibitem{Eisebitt:2003}
S.~Eisebitt et~al.
\newblock Polarization effects in coherent scattering from magnetic specimen:
  Implications for x-ray holography, lensless imaging, and correlation
  spectroscopy.
\newblock {\em Physical Review B}, 68:104419, Jan 2003.

\bibitem{Pierce:2007}
M.~S. Pierce, C.~Buechler, L.~Sorensen, and S.~Kevan.
\newblock Disorder-induced magnetic memory: Experiments and theories.
\newblock {\em Physical Review B}, 75:144406, Jan 2007.

\bibitem{Elser:2010}
V.~Elser and S.~Eisebitt.
\newblock Uniqueness transition in noisy phase retrieval.
\newblock {\em New Journal of Physics}, 2010.
\newblock Submitted.

\bibitem{Hannon:1988}
J.~P. Hannon, G.~T. Trammell, M.~Blume, and D.~Gibbs.
\newblock X-ray resonance exchange scattering.
\newblock {\em Physical Review Letters}, 61(10):1245, Sep 1988.

\bibitem{Kortright:2001}
J.~B. Kortright et~al.
\newblock Soft-x-ray small-angle scattering as a sensitive probe of magnetic
  and charge heterogenity.
\newblock {\em Physical Review B}, 64:092401, Jan 2001.

\bibitem{Elser:2007}
V.~Elser, I.~Rankenburg, and P.~Thibault.
\newblock Searching with iterated maps.
\newblock {\em Proceedings of the National Academy of Sciences},
  104(2):418--423, Jan 2007.

\bibitem{Elser:2003}
V.~Elser.
\newblock Phase retrieval by iterated projections.
\newblock {\em Journal of the Optical Society of America A}, 20(1):40--55,
  2003.

\bibitem{Luke:2005}
D.~R. Luke.
\newblock Relaxed averaged alternating reflections for diffraction imaging.
\newblock {\em Inverse problems}, 21(1):37--50, 2005.

\bibitem{Thibault:Thesis}
P.~Thibault.
\newblock {\em Algorithmic methods in diffraction microscopy}.
\newblock PhD thesis, Cornell University, 2007.

\bibitem{Durr:1997}
H.~D\"{u}rr et~al.
\newblock Element-specific magnetic anisotropy determined by transverse
  magnetic circular x-ray dichroism.
\newblock {\em Science}, 277:213--215, Jan 1997.

\bibitem{Hubert}
A.~Hubert and R.~Sch\"{a}fer.
\newblock {\em Magnetic Domains: The analysis of magnetic microstructures}.
\newblock Springer-Verlag, Berlin Heidelberg, 1998.

\bibitem{Schlotter:2006}
W.~Schlotter et~al.
\newblock Multiple reference fourier transform holography with soft x-rays.
\newblock {\em Applied Physics}, 89(163112):1--3, Jan 2006.

\end{thebibliography}

\end{document}